\documentclass[twocolumn,eqsecnum,nofootinbib,showpacs,preprintnumbers,widetext]{revtex4}

\usepackage{latexsym}
\usepackage{amsmath,amsfonts}

\usepackage{amsmath,amsfonts,latexsym,amssymb,times}
\allowdisplaybreaks[4]

\newcommand{\rf}[1]{(\ref{#1})}

\def\be{\begin{equation}}
\def\ee{\end{equation}}
\def\beq{\begin{eqnarray}}
\def\eeq{\end{eqnarray}}

\def\parline{\,\partial\kern -0.55em /\,\,}

\def\half{{\frac{1}{2}}}

\def\LL{{\cal L}}

\def\TT{{\cal T}}

\def\Obf{{\bf O}}

\def\phik{|\phi\rangle}

\def\xik{|\xi\rangle}

\def\smzero{{\scriptscriptstyle (0)}}
\def\smone{{\scriptscriptstyle (1)}}

\def\smzero{{\scriptscriptstyle (0)}}
\def\smone{{\scriptscriptstyle (1)}}

\def\mubf{{\boldsymbol{\mu}}}

\def\smponetwo{{\scriptscriptstyle [1,2]}}

\def\Cwt{\widetilde{C}}

\def\alpar{\alpha\partial}
\def\albpar{\bar\alpha\partial}

\def\AdS{{\rm AdS}}
\def\CFT{{\rm CFT}}
\def\sc{{\rm sc}}
\def\cur{{\rm cur}}
\def\sh{{\rm sh}}
\def\eff{{\rm eff}}

\def\norm{{\rm norm}}
\def\stand{{\rm stand}}
\def\impr{{\rm impr}}

\def\Cb{\bar{C}}
\def\eb{\bar{e}}

\begin{document}

\preprint{FIAN-TD-2011-12; arXiv: 1110.3749 [hep-th] }

\title{ Anomalous conformal currents, shadow fields and massive AdS fields }

\author{ R.R. Metsaev}

\email{metsaev@lpi.ru}

\affiliation{ Department of Theoretical Physics, P.N. Lebedev Physical
Institute, Leninsky prospect 53,  Moscow 119991, Russia}

\begin{abstract}
In the framework of gauge invariant approach involving Stueckelberg and auxiliary fields,
totally symmetric arbitrary spin anomalous conformal current and shadow field in flat
space-time of dimension greater than or equal to four are studied. Gauge invariant
differential constraints for such anomalous conformal current and shadow field and
realization of global conformal symmetries are obtained. Gauge invariant two-point vertex of
the arbitrary spin anomalous shadow field is also obtained. In Stueckelberg gauge frame, the
two-point gauge invariant vertex becomes the standard two-point vertex of CFT. Light-cone
gauge two-point vertex of the arbitrary spin anomalous shadow field is derived. The AdS/CFT
correspondence for arbitrary spin anomalous conformal current and shadow field and the
respective normalizable and non-normalizable modes of massive arbitrary spin AdS field is
studied. The AdS field is considered in modified de Donder gauge which simplifies
considerably the study of AdS/CFT correspondence. We show that on-shell leftover gauge
symmetries of bulk massive field are related to gauge symmetries of boundary anomalous
conformal current and shadow field, while the modified de Donder gauge condition for bulk
massive field is related to differential constraints for boundary anomalous conformal current
and shadow field.
\end{abstract}


\maketitle

\section{Introduction}

We start with a brief recall of some basic notions of CFT. In space-time of dimension $d\geq
4$, fields of CFT can be separated into two groups: conformal currents and shadow fields.
This is to say that field having Lorentz algebra spin $s$, $s\geq 1$, and conformal dimension
$\Delta = s+d-2$ is referred to as conformal current with canonical dimension, while field
having Lorentz algebra spin $s$, $s\geq 1$, and conformal dimension $\Delta > s+d-2$ is
referred to as anomalous conformal current. Accordingly, field having Lorentz algebra spin
$s$, $s\geq 1$, and conformal dimension
$\Delta = 2 - s$ is referred to as shadow field with canonical dimension%
\footnote{ Shadow fields having the canonical dimension are used to build
conformal invariant equations of motion and Lagrangian formulations for
conformal fields in Refs.\cite{Fradkin:1985am}. Interesting discussion of
shadow field dualities may be found in Ref.\cite{Petkou:1994ad}.},
while field having Lorentz algebra spin $s$, $s\geq 1$, and conformal
dimension $\Delta < 2 - s$ is referred to as anomalous shadow field.

In Refs.\cite{Metsaev:2008fs,Metsaev:2009ym}, we developed the gauge invariant formulation of
the conformal currents and shadow fields having the canonical conformal dimensions. We recall
that, in the framework of AdS/CFT correspondence, such conformal currents and shadow fields
are related to massless AdS fields. In Ref.\cite{Metsaev:2010zu}, we extended our approach to
the case of low spin-$s$, $s=1,2$, anomalous conformal currents and shadow fields. The
purpose of this paper is to develop gauge invariant approach to bosonic totally symmetric
arbitrary spin-$s$ anomalous conformal current and shadow field. In the framework of AdS/CFT
correspondence, spin-$s$ anomalous conformal current and shadow field are related to spin-$s$
massive AdS field. Massive totally symmetric spin-$s$ AdS fields with even $s\geq 4 $ form
leading Regge trajectory of AdS string theory. Therefore, extension of our approach to the
case of arbitrary $s$ is important. Our approach to anomalous conformal current and shadow
field is summarized as follows.
\\
{\bf i}) Starting with the field content of anomalous conformal current (and anomalous shadow
field) in the standard CFT, we introduce Stueckelberg fields and auxiliary fields. In other
words, we extend space of fields entering the standard CFT.
\\
{\bf ii}) On the extended space of fields entering our approach, we introduce
differential constraints, gauge transformations, and conformal algebra
transformations. The differential constraints are required to be invariant
under the gauge transformations and the conformal algebra transformations.
\\
{\bf iii})  The gauge symmetries and the differential constraints allow us to match our
approach and the standard CFT. Namely, by imposing gauge conditions to exclude the
Stueckelberg fields and by solving differential constraints to exclude the auxiliary fields,
we obtain formulation of anomalous conformal current and shadow field in the standard CFT.

Besides the gauge invariant approach, we discuss the anomalous conformal
current and shadow field by using Stueckelberg gauge and light-cone gauge
conditions. Reasons for discussing these two gauge conditions are as follows.

\noindent {\bf i}) The Stueckelberg gauge reduces our approach to the standard formulation of
CFT. This is to say that use of the Stueckelberg gauge allows us to demonstrate how the
standard approach to anomalous conformal current and shadow field is connected with our gauge
invariant approach.

\noindent {\bf ii}) Studying of CFT in the light-cone gauge frame is motivated by the
conjectured duality of the supersymmetric Yang-Mills theory and AdS superstring theory
\cite{Maldacena:1997re}. We expect, by analogy with flat space, that a quantization of the
type IIB Green-Schwarz AdS superstring \cite{Metsaev:1998it} will be straightforward only in
the light-cone gauge \cite{Metsaev:2000yf,Metsaev:2000yu,Metsaev:2000mv}. Therefore we think
that, from the stringy perspective of AdS/CFT correspondence, the light-cone approach to CFT
deserves to be understood better. In this respect, we note that our approach provides quick
access to the light-cone gauge formulation of CFT. This implies that our approach gives easy
access to the studying of AdS/CFT correspondence in light-cone gauge. This seems to be
important for future application of our approach to the studying of string/gauge theory
duality.

We use our gauge invariant formulation of CFT for the studying AdS/CFT correspondence between
arbitrary spin massive AdS field and the corresponding arbitrary spin boundary anomalous
conformal current and shadow field. Namely, we show that non-normalizable modes of arbitrary
spin-$s$ massive AdS field are related to arbitrary spin-$s$ anomalous shadow field, while
normalizable modes of arbitrary spin-$s$ massive AdS field are related to arbitrary spin-$s$
anomalous conformal current. We recall that, in earlier literature, the AdS/CFT
correspondence between non-normalizable modes of massive spin-1 and spin-2 AdS fields and the
corresponding spin-1 and spin-2 anomalous shadow fields was studied in
Refs.\cite{Mueck:1998iz,Polishchuk:1999nh}. The AdS/CFT correspondence for spin-$s$ massive
AdS field with $s>2$ and the corresponding spin-$s$ anomalous conformal current and shadow
field has not been considered in the earlier literature. Our treatment of AdS/CFT
correspondence is summarized as follows.
\\
{\bf i}) We exploit the CFT adapted gauge invariant approach to massive AdS
fields and modified de Donder gauge obtained in
Refs.\cite{Metsaev:2008ks,Metsaev:2009hp}. The modified de Donder gauge leads
to the simple decoupled bulk equations of motion which are easily solved. We
show that the two-point gauge invariant vertex of the arbitrary spin-$s$
anomalous shadow field does indeed emerge from massive arbitrary spin-$s$ AdS
field action when it is evaluated on solution of the Dirichlet problem.
Throughout this paper the AdS field action evaluated on the solution of the
Dirichlet problem is referred to as effective action.
\\
{\bf ii}) The number of boundary gauge fields involved in our
approach to the anomalous conformal current (or anomalous shadow field)
coincides with the number of gauge fields involved in the CFT
adapted formulation of massive AdS field in Ref.\cite{Metsaev:2009hp}.
\\
{\bf iii}) The number of gauge transformation parameters involved in our
approach to the anomalous conformal current (or anomalous shadow field)
coincides with the number of gauge transformation parameters involved in the
CFT adapted gauge invariant formulation of massive AdS field in
Ref.\cite{Metsaev:2009hp}.
\\
{\bf iv}) The modified de Donder gauge for massive bulk field is related to the differential
constraint for boundary anomalous conformal current (or anomalous shadow field).
\\
{\bf v}) On-shell leftover gauge symmetries of massive bulk field are related to the gauge
symmetries of boundary anomalous conformal current (or anomalous shadow field).

Our paper is organized as follows.

In Sec. \ref{sec02}, we summarize our notation and conventions.

Section \ref{sec03} is devoted to gauge invariant formulation of arbitrary spin-$s$ anomalous
conformal current. We discuss gauge symmetries and realization of global conformal algebras
symmetries on space of gauge fields we use for the description of the anomalous conformal
current. We demonstrate how our gauge invariant approach is related to the standard CFT.
Also, using our approach, we obtain light-cone gauge description of the arbitrary spin-$s$
anomalous conformal current.

In Sec. \ref{sec04}, we extend results in Sec. \ref{sec03} to the case of arbitrary spin-$s$
anomalous shadow field.  Also, we find gauge invariant two-point vertex for the arbitrary
spin-$s$ anomalous shadow field. We discuss the two-point vertex in Stueckelberg gauge frame
and in light-cone gauge frame.

In Sec. \ref{sec09}, we discuss the two-point current-shadow field
interaction vertex.

In Sec. \ref{secAdS/CFT}, we review the CFT adapted gauge invariant approach to massive
arbitrary spin AdS field. Because the use of the modified de Donder gauge makes our study of
AdS/CFT correspondence for arbitrary spin fields similar to the one for scalar field, we
briefly review the AdS/CFT correspondence for the scalar field.

Section \ref{secAdS/CFTcur} is devoted to the study of AdS/CFT correspondence between
normalizable modes of massive spin-$s$ AdS field and spin-$s$ anomalous conformal current,
while, in Sec.\ref{secAdS/CFTsh}, we study the AdS/CFT correspondence between
non-normalizable modes of massive spin-$s$ AdS field and spin-$s$ anomalous shadow field.

Section \ref{conlus} summarizes our conclusions and suggests directions for
future research.

In Appendix, we present some details of matching of the bulk and boundary
conformal boost symmetries.

\section{Preliminaries}
\label{sec02}

\subsection{Notation}

We use the following conventions. The Cartesian coordinates in
$d$-dimensional flat space-time are denoted by  $x^a$, while derivatives with
respect to $x^a$ are denoted by $\partial_a$, $\partial_a \equiv
\partial/\partial x^a$. The vector indices of the Lorentz algebra $so(d-1,1)$ take
the values $a,b,c,e=0,1,\ldots ,d-1$. We use the mostly positive flat metric
tensor $\eta^{ab}$ and, to simplify our expressions, we drop $\eta_{ab}$ in
scalar products: $X^aY^a \equiv \eta_{ab}X^a Y^b$. Creation operators
$\alpha^a$, $\alpha^z$, $\zeta$ and the respective annihilation operators
$\bar{\alpha}^a$,
$\bar{\alpha}^z$, $\bar\zeta$ are referred to as oscillators.%
\footnote{ We use oscillators to handle the many indices appearing for tensor
fields (discussion of oscillator formulation may be found in
Refs.\cite{Bekaert:2006ix,Boulanger:2008up}.)}
Commutation relations of the oscillators, the vacuum $|0\rangle$, and
hermitian conjugation rules are defined as
\beq
\label{man01-13102011-07} && [\bar{\alpha}^a,\alpha^b]=\eta^{ab}\,, \qquad
[\bar\alpha^z,\alpha^z]=1\,,\qquad [\bar\zeta,\zeta]=1\,,\qquad
\\
\label{man01-13102011-08} && \bar\alpha^a |0\rangle = 0\,,\qquad \bar\alpha^z
|0\rangle = 0\,,\qquad \bar\zeta |0\rangle = 0\,,
\\
\label{man01-13102011-09} && \alpha^{a\dagger} = \bar\alpha^a\,, \qquad \
\alpha^{z\dagger} = \bar\alpha^z \,, \qquad \ \zeta^\dagger = \bar\zeta\,.
\eeq
The oscillators $\alpha^a$, $\bar\alpha^a$ and $\alpha^z$, $\zeta$, $\bar\alpha^z$,
$\bar\zeta$ transform in the respective vector and scalar representations of the Lorentz
algebra $so(d-1,1)$. Throughout this paper we use operators constructed out of the
derivatives, coordinates, and the oscillators,
\beq
&& \Box \equiv \partial^a\partial^a\,,\qquad \ \
x\partial \equiv x^a \partial^a \,, \qquad x^2 \equiv x^a x^a\,, \qquad
\\
&& \alpha\partial \equiv \alpha^a\partial^a\,,\qquad\albpar
\equiv \bar\alpha^a\partial^a\,,
\\
&& \alpha^2 \equiv \alpha^a\alpha^a\,,\qquad \ \bar\alpha^2 \equiv
\bar\alpha^a\bar\alpha^a\,,
\\
&& N_\alpha \equiv \alpha^a \bar\alpha^a  \,,
\qquad
N_z \equiv \alpha^z \bar\alpha^z \,,
\qquad
N_\zeta \equiv \zeta \bar\zeta \,, \qquad
\\
\label{mabdef0001} && M^{ab} \equiv \alpha^a \bar\alpha^b -
\alpha^b\bar\alpha^a\,,
\\
\label{man01-13102011-10} && \Pi^\smponetwo \equiv 1
-\alpha^2\frac{1}{2(2N_\alpha +d)}\bar\alpha^2\,,
\\
\label{man01-16102011-04} && \mubf \equiv 1 -
\frac{1}{4}\alpha^2\bar\alpha^2\,,
\\
\label{man01-16102011-05}
&& \Cwt^a \equiv \alpha^a - \alpha^2 \frac{1}{2N_\alpha + d -2}\bar\alpha^a\,,
\\
\label{man01-16102011-06} && \Cb_\perp^a \equiv \bar\alpha^a - \frac{1}{2}
\alpha^a \bar\alpha^2 \,,
\eeq

\begin{widetext}

\beq
\label{man01-13102011-11} && r_\zeta \equiv \left(\frac{(s+\frac{d-4}{2}
-N_\zeta)(\kappa - s-\frac{d-4}{2} + N_\zeta)(\kappa + 1 +
N_\zeta)}{2(s+\frac{d-4}{2}-N_\zeta - N_z)(\kappa +N_\zeta -N_z) (\kappa+
N_\zeta - N_z +1)}\right)^{1/2}\,,
\nonumber\\[-7pt]
&&
\\[-7pt]
&& r_z \equiv  \left(\frac{(s+\frac{d-4}{2} -N_z)(\kappa + s + \frac{d-4}{2}
- N_z)(\kappa - 1 - N_z)}{2(s+\frac{d-4}{2}-N_\zeta-N_z)(\kappa + N_\zeta -
N_z) (\kappa +N_\zeta - N_z -1)}\right)^{1/2}\,,
\nonumber
\eeq
where parameter $\kappa$ appearing in \rf{man01-13102011-11} is defined below
in \rf{man01-11102011-04}. Throughout the paper the notation $\lambda \in
[n]_2$ implies that $\lambda =-n,-n+2,-n+4,\ldots,n-4, n-2,n$:
\be \label{man01-16102011-02}
\lambda \in [n]_2 \ \ \Longrightarrow \ \ \lambda
=-n,-n+2,-n+4,\ldots,n-4, n-2,n\,.
\ee
Often, we use the following set of scalar, vector, and totally symmetric
tensor fields of the Lorentz algebra $so(d-1,1)$:
\be \label{man01-16102011-07}
\phi_\lambda^{a_1\ldots a_{s'}}\,,
\qquad
s'=0,1,\ldots, s\,,
\qquad
\lambda \in [s-s']_2\,,
\ee
where $\phi_\lambda^{a_1\ldots a_{s'}}$ is rank-$s'$ totally symmetric traceful tensor field
of the Lorentz algebra $so(d-1,1)$. To illustrate the field content given in
\rf{man01-16102011-07}, we use shortcut $\phi_\lambda^{s'}$ for the field
$\phi_\lambda^{a_1\ldots a_{s'}}$ and note that fields in \rf{man01-16102011-07} can be
represented as
\be  \label{man01-16102011-08}
\begin{array}{ccccccccc}
&  & &  & \phi_0^s& & & &
\\[12pt]
&  & & \phi_{-1}^{s-1}  & & \phi_1^{s-1} & & &
\\[12pt]
& & \ldots  &   & \ldots & & \ldots  &  &
\\[12pt]
& \phi_{1-s}^1 & & \phi_{3-s}^1 & \ldots & \phi_{s-3}^1 & &
\phi_{s-1}^1 &
\\[12pt]
\phi_{-s}^0 & & \phi_{2-s}^0 & & \ldots & & \phi_{s-2}^0 & &
\phi_s^0
\end{array}
\ee

\end{widetext}

Our conventions for light-cone frame are as follows. The space-time
coordinates are decomposed as $x^a= x^+, x^-,x^i$, where the coordinates in
$\pm$ directions are defined as $x^\pm=(x^{d-1} \pm x^0)/\sqrt{2}$ and $x^+$
is taken to be a light-cone time. Vector indices of the $so(d-2)$ algebra take
values $i,j =1,\ldots, d-2$. We use the following conventions for the
derivatives:
\be \partial^i=\partial_i\equiv\partial/\partial x^i\,, \qquad
\partial^\pm=\partial_\mp \equiv
\partial/\partial x^\mp \,.\ee

\subsection{Global conformal symmetries }

In $d$-dimensional flat space-time, the conformal algebra $so(d,2)$ consists
of translation generators $P^a$, dilatation generator $D$, conformal boost
generators $K^a$, and generators of the $so(d-1,1)$ Lorentz algebra $J^{ab}$.
We use the following nontrivial commutators of the conformal algebra:
\beq
&& {}[D,P^a]=-P^a\,, \hspace{0.5cm}  {}[P^a,J^{bc}]=\eta^{ab}P^c
-\eta^{ac}P^b, \ \ \ \
\nonumber\\
&& [D,K^a]=K^a\,, \hspace{0.7cm}
[K^a,J^{bc}]=\eta^{ab}K^c - \eta^{ac}K^b,\qquad
\nonumber\\[-10pt]
\label{man01-12112010-01} &&
\\[-10pt]
&& \hspace{1.5cm} {}[P^a,K^b]=\eta^{ab}D - J^{ab}\,,
\nonumber\\
&& \hspace{1.5cm} [J^{ab},J^{ce}]=\eta^{bc}J^{ae}+3\hbox{ terms} \,.
\nonumber
\eeq

Let $\phi$ denotes anomalous conformal current (or anomalous shadow field) in
the $d$-dimensional flat space-time. Under the action of conformal algebra,
the $\phi$ transforms as
\be \label{04092008-01} \delta_{\hat{G}} \phi  = \hat{G} \phi \,, \ee
where the realization of the conformal algebra generators $\hat{G}$ on space
of $\phi$ is given by
\beq
\label{conalggenlis01} && P^a = \partial^a \,,
\\
\label{conalggenlis02} && J^{ab} = x^a\partial^b -  x^b\partial^a + M^{ab}\,,
\\
\label{conalggenlis03} && D = x\partial  + \Delta\,,
\\
\label{conalggenlis04} && K^a = K_{\Delta,M}^a + R^a\,,
\\
\label{kdelmdef01} && \hspace{1cm} K_{\Delta,M}^a \equiv
-\frac{1}{2}x^2\partial^a + x^a D + M^{ab}x^b\,.\qquad
\eeq
In relations \rf{conalggenlis02}-\rf{conalggenlis04}, $\Delta$ is an operator
of conformal dimension, while $M^{ab}$ is a spin operator of the Lorentz
algebra,
\be  [M^{ab},M^{ce}]=\eta^{bc}M^{ae}+3\hbox{ terms} \,. \ee
The spin operator of the Lorentz algebra is well known for arbitrary spin anomalous conformal
current and shadow field (see \rf{mabdef0001}). In general, operator $R^a$ appearing in
\rf{conalggenlis04} depends on the
derivatives and does not depend on the space-time coordinates.%
\footnote{ For the anomalous conformal currents and shadow fields considered in this
paper, the operator $R^a$ is independent of the derivatives. Dependence of
the operator $R^a$ on the derivatives appears in the ordinary-derivative
approach to conformal fields (see
Refs.\cite{Metsaev:2010kp}-\cite{Metsaev:2007fq}).}
In the standard CFT, the operator $R^a$ is equal to zero, while, in the gauge
invariant approach to anomalous conformal current and shadow field we develop
in this paper, the operator $R^a$ is nontrivial. This is to say that, in the
framework of our gauge invariant approach, the complete description of the
conformal current and shadow field requires, among other things, finding the
operator $R^a$.

\section{ Arbitrary spin anomalous conformal current}
\label{sec03}

\subsection{ Gauge invariant formulation }
\label{sec03G}

{\bf Field content}. To develop gauge invariant formulation of arbitrary
spin-$s$ anomalous conformal current in flat space of dimension $d\geq 4$ we
use the following fields:
\be \label{man01-10112011-01}
\phi_{\cur,\lambda}^{a_1\ldots a_{s'}}\,,
\qquad
s'=0,1,\ldots, s\,,
\qquad
\lambda \in [s-s']_2\,.
\ee
We note that:\\
{\bf i}) In \rf{man01-10112011-01}, the fields $\phi_{\cur,\lambda}$ and
$\phi_{\cur,\lambda}^a$ are the respective scalar and vector fields of the Lorentz algebra,
while the field $\phi_{\cur,\lambda}^{a_1\ldots a_{s'}}$, $s'>1$, is rank-$s'$ totally
symmetric traceful tensor field of the Lorentz algebra $so(d-1,1)$. Using shortcut
$\phi_\lambda^{s'}$ for the field $\phi_{\cur,\lambda}^{a_1\ldots a_{s'}}$, fields in
\rf{man01-10112011-01} can be represented as in \rf{man01-16102011-08}.
\\
{\bf ii}) The tensor fields $\phi_{\cur,\,\lambda}^{a_1\ldots a_{s'}}$ with
$s'\geq 4 $ satisfy the double-tracelessness constraint
\be \label{man01-10112011-02} \phi_{\cur,\lambda}^{aabba_5\ldots a_{s'}}=0\,,
\qquad s'=4,5,\ldots, s\,. \ee
\\
{\bf iii}) The fields $\phi_{\cur,\lambda}^{a_1\ldots a_{s'}}$ have the
following conformal dimensions:
\be \label{man01-10112011-03}
\Delta(\phi_{\cur,\lambda}^{a_1\ldots a_{s'}})= \frac{d}{2} + \kappa +
\lambda\,.
\ee
{\bf iv)} In the framework of AdS/CFT correspondence, $\kappa$ is related to
the mass parameter $m$ of spin-$s$ massive field in $AdS_{d+1}$ as
\be \label{man01-11102011-04} \kappa \equiv \sqrt{m^2 + \Bigl( s +
\frac{d-4}{2}\Bigr)^2 }\,.\ee

In order to obtain the gauge invariant description in an easy--to--use form
we use the oscillators and introduce a ket-vector $|\phi_\cur\rangle$
defined by
\beq \label{man01-10112011-05}
&& |\phi_\cur\rangle = \sum_{s'=0}^s |\phi_\cur^{s'}\rangle \,,
\\
&& |\phi_\cur^{s'}\rangle
\nonumber\\
&&  = \!\!\! \sum_{\lambda \in [s-s']_2}\!\!
\frac{\zeta_{\phantom{z}}^{\frac{s-s'+\lambda}{2}}
\alpha_z^{\frac{s-s'-\lambda}{2}}\alpha^{a_1}\ldots
\alpha^{a_{s'}}}{s'!\sqrt{(\frac{s-s'+\lambda}{2})!
(\frac{s-s'-\lambda}{2})!}} \, \phi_{\cur,\lambda}^{a_1\ldots a_{s'}}
|0\rangle\,.
\nonumber
\eeq
From \rf{man01-10112011-05}, we see that the ket-vector $|\phi_\cur\rangle$
is degree-$s$ homogeneous polynomial in the oscillators $\alpha^a$,
$\alpha^z$, $\zeta$, while the ket-vector $|\phi_\cur^{s'}\rangle$ is
degree-$s'$ homogeneous polynomial in the oscillators $\alpha^a$, i.e., these
ket-vectors satisfy the relations
\beq \label{man01-10112011-06}
&& (N_\alpha + N_z + N_\zeta -s) |\phi_\cur\rangle =0 \,,
\\
\label{man01-10112011-07} && (N_\alpha -s') |\phi_\cur^{s'}\rangle =0\,.
\eeq
In terms of the ket-vector $|\phi_\cur\rangle$, double-tracelessness
constraint \rf{man01-10112011-02} takes the form%
\footnote{ In this paper we adapt the formulation in terms of the double tracelless gauge
fields \cite{Fronsdal:1978vb}. To develop the gauge invariant approach one can use
unconstrained gauge fields studied in Refs.\cite{Francia:2002aa}. Discussion of other gauge
fields which seem to be most suitable for the theory of interacting fields may be found,
e.g., in \cite{Alkalaev:2003qv}.}
\beq \label{man01-10112011-08}
&& (\bar{\alpha}^2)^2 |\phi_\cur\rangle  = 0 \,.
\eeq

{\bf Differential constraint}. We find the following differential constraint
for the anomalous conformal current:
\beq \label{man01-10112011-09}
&& \Cb_\cur |\phi_\cur\rangle = 0 \,,
\\
\label{man01-10112011-10} && \Cb_\cur  = \albpar - \half \alpar \bar\alpha^2
- \eb_{1,\cur} \Pi^\smponetwo + \half e_{1,\cur} \bar\alpha^2 \,,\qquad\quad
\\
\label{man01-10112011-12} && e_{1,\cur} =  \zeta r_\zeta \Box + \alpha^zr_z
\,,
\nonumber\\[-10pt]
&&
\\[-10pt]
&& \eb_{1,\cur} = -r_\zeta\bar\zeta  - r_z \bar\alpha^z \Box \,,
\nonumber
\eeq
where the operators $\Pi^\smponetwo$, $r_\zeta$, $r_z$ are defined in
\rf{man01-13102011-10}, \rf{man01-13102011-11}. One can make sure that
constraint \rf{man01-10112011-09} is invariant under gauge transformation and
conformal algebra transformations which we discuss below.

{\bf Gauge symmetries}. We now discuss gauge symmetries of the
anomalous conformal current. To this end we introduce the following gauge
transformation parameters:
\be \label{man01-10112011-15}
\xi_{\cur,\lambda}^{a_1\ldots a_{s'}}\,,
\qquad s'=0,1,\ldots, s-1\,,
\qquad
\lambda \in [s-1-s']_2\,.
\ee
We note that\\
{\bf i}) In \rf{man01-10112011-15}, the gauge transformation parameters $\xi_{\cur,\lambda}$
and $\xi_{\cur,\lambda}^a$ are the respective scalar and vector fields of the Lorentz
algebra, while the gauge transformation parameter $\xi_{\cur,\lambda}^{a_1\ldots a_{s'}}$,
$s'>1$, is rank-$s'$ totally symmetric tensor field of the Lorentz algebra $so(d-1,1)$.
\\
{\bf ii}) The gauge transformation parameters $\xi_{\cur,\lambda}^{a_1\ldots
a_{s'}}$ with $s'\geq 2 $ satisfy the tracelessness constraint,
\be \label{man01-10112011-16} \xi_{\cur,\lambda}^{aaa_3\ldots a_{s'}}=0\,,
\qquad s'= 2,3,\ldots, s-1\,. \ee
{\bf iii}) The gauge transformation parameters $\xi_{\cur,
\lambda}^{a_1\ldots a_{s'}}$ have the conformal dimensions
\be \label{man01-10112011-17}
\Delta(\xi_{\cur,\lambda}^{a_1\ldots a_{s'}})= \frac{d}{2} + \kappa + \lambda
-1\,.
\ee

Now, as usually, we collect the gauge transformation parameters in a ket-vector
$|\xi_\cur\rangle$ defined by
\beq \label{man01-10112011-18}
&& |\xi_\cur\rangle  = \sum_{s'=0}^{s-1} |\xi_\cur^{s'}\rangle\,,
\\
&& |\xi_\cur^{s'}\rangle
\nonumber\\
&& =\!\!\!\!\sum_{\lambda \in [s-1-s']_2}\!\!\!
\frac{\zeta_{\phantom{z}}^{\frac{s-1-s'+\lambda}{2}}
\alpha_z^{\frac{s-1-s'-\lambda}{2}}\alpha^{a_1}\ldots
\alpha^{a_{s'}}}{s'!\sqrt{(\frac{s-1-s'+\lambda}{2})!
(\frac{s-1-s'-\lambda}{2})!}} \, \xi_{\cur,\lambda}^{a_1\ldots a_{s'}}
|0\rangle\,.
\nonumber
\eeq
The ket-vectors $|\xi_\cur\rangle$, $|\xi_\cur^{s'}\rangle$ satisfy the
algebraic constraints
\beq \label{man01-10112011-19}
&& (N_\alpha + N_z + N_\zeta - s +1 ) |\xi_\cur\rangle = 0 \,,
\\
\label{man01-10112011-20} && (N_\alpha -s')|\xi_\cur^{s'}\rangle  = 0\,,
\eeq
which tell us that $|\xi_\cur\rangle$ is a degree-$(s-1)$ homogeneous
polynomial in the oscillators $\alpha^a$, $\alpha^z$, $\zeta$, while the
ket-vector $|\xi_\cur^{s'}\rangle$ is degree-$s'$ homogeneous polynomial in
the oscillators $\alpha^a$. In terms of the ket-vector $|\xi_\cur\rangle$,
tracelessness constraint \rf{man01-10112011-16} takes the form
\be \label{man01-10112011-21} \bar\alpha^2 |\xi_\cur\rangle = 0 \,.\ee

Gauge transformation can entirely be written in terms of $|\phi_\cur\rangle$
and $|\xi_\cur\rangle$. This is to say that gauge transformation takes the
form
\beq
\label{man01-10112011-22} && \delta |\phi_\cur\rangle = G_\cur
|\xi_\cur\rangle\,,
\\
&& G_\cur  \equiv \alpar - e_{1,\cur} - \alpha^2 \frac{1}{2N_\alpha + d- 2}
\eb_{1,\cur} \,,\qquad
\eeq
where $e_{1,\cur}$, $\eb_{1,\cur}$ are given in \rf{man01-10112011-12}. As we
have already said, constraint \rf{man01-10112011-09} is invariant under gauge
transformation \rf{man01-10112011-22}.

{\bf Realization of conformal algebra symmetries}. To complete the gauge invariant
formulation of the spin-$s$ anomalous conformal current we provide realization of the
conformal algebra symmetries on space of the ket-vector $|\phi_\cur\rangle$. All that is
needed is to fix the operators $M^{ab}$, $\Delta$, and $R^a$ and insert then these operators
into \rf{conalggenlis01}-\rf{conalggenlis04}. Realization of the spin operator $M^{ab}$ on
ket-vector $|\phi_\cur\rangle$ \rf{man01-10112011-05} is given in \rf{mabdef0001}, while
realization of the operator $\Delta$,
\beq \label{man01-15102011-36a1}
\Delta_\cur & = & \frac{d}{2} + \nu\,,
\qquad \nu \equiv \kappa + N_\zeta - N_z \,,
\eeq
can be read from \rf{man01-10112011-03}. In the gauge invariant formulation,
finding the operator $R^a$ provides the real difficulty. We find the
following realization of the operator $R^a$ on space of $|\phi_\cur\rangle$:
\beq  \label{man01-15102011-43}
R_\cur^a  & = & - 2 \zeta r_\zeta \Bigl( (\nu+1) \bar\alpha^a  -
\bar{C}_\perp^a\Bigr)
\nonumber\\
& - & 2\Bigl( \nu \Cwt^a + \alpha^2 \frac{1}{2N_\alpha+d-2} \Cb_\perp^a
\Bigr)r_z\bar\alpha^z\,,\qquad
\eeq
where the operators  $\Cwt^a$, $\Cb_\perp^a$ are given in
\rf{man01-16102011-05},\rf{man01-16102011-06}, while the operators $r_\zeta$, $r_z$ are
defined in \rf{man01-13102011-11}.

We derived the differential constraint, gauge transformation, and the realization of the
operator $R_\cur^a$ by generalizing our results for the spin-$s$ conformal current with the
canonical dimension which we obtained in Ref.\cite{Metsaev:2008fs}. It is worthwhile to note
that results in this section can also be obtained by using the tractor approach in
Refs.\cite{Gover:2008sw,Gover:2008pt}%
\footnote{ In mathematical literature, interesting discussion of the tractor approach may
be found in Ref.\cite{Gov}.}.
Our constraint \rf{man01-10112011-09} and gauge transformation \rf{man01-10112011-22} can be
matched with the ones in Ref.\cite{Gover:2008sw} by using appropriate field redefinitions.
The basis of the fields we use in this paper turns out to be more convenient for the study of
AdS/CFT correspondence.  To summarize, our fields \rf{man01-10112011-01} can be written as a
tractor rank-$s$ tensor field subject to a Thomas-D divergence type constraint in
Ref.\cite{Gover:2008sw}. A similar construction was used to describe spin-$s$ massive bulk
field in Ref.\cite{Gover:2008sw}. Note that, in our approach, we use our fields
\rf{man01-10112011-01} for the discussion of spin-$s$ anomalous conformal current.

\subsection{ Stueckelberg gauge frame }
\label{sec03S}

We proceed with discussion of the spin-$s$ anomalous conformal current in the Stueckelberg
gauge frame. To this end we note that the Stueckelberg gauge frame is achieved through the
use of differential constraint \rf{man01-10112011-09} and the Stueckelberg gauge condition.
From \rf{man01-10112011-22}, we see that a field defined by
\be \bar\alpha^z \Pi^\smponetwo |\phi_\cur\rangle \ee
transforms as Stueckelberg field. Therefore this field can be gauged away via Stueckelberg
gauge fixing,
\be \label{man01-29012012-01} \bar\alpha^z \Pi^\smponetwo |\phi_\cur\rangle = 0\,, \ee
where $\Pi^\smponetwo$ is given in \rf{man01-13102011-10}. Using gauge condition
\rf{man01-29012012-01} and differential constraint \rf{man01-10112011-09}, we find the
relations
\beq  \label{10112011-04}
&& \hspace{-0.9cm}  \bar\alpha^2 |\phi_\cur^{s'}\rangle =0\,,
\\
\label{10112011-05} && \hspace{-0.9cm} |\phi_\cur^{s'}\rangle = X_\cur
\zeta^{s-s'} (\albpar)^{s-s'} |\phi_\cur^s\rangle\,,
\\[5pt]
&& \hspace{-0.9cm} X_\cur \!\equiv\!  \frac{(-)^{s-s'}}{(s-s')!} \Bigl(
\frac{2^{s-s'} \Gamma(\kappa\! -\! s\! -\frac{d-4}{2}\!) \Gamma(\kappa\!+\!
s\!-\! s')}{\Gamma(\kappa\! - \! s'\!-\!\frac{d-4}{2})\Gamma(\kappa)}\!
\Bigr)^{1/2}.
\nonumber
\eeq
Relation \rf{10112011-04} tells us that all fields $\phi_{\cur,\lambda}^{a_1\ldots a_{s'}}$
with $s'\geq 2$ become traceless. From relation \rf{10112011-05}, we learn that the fields
$\phi_{\cur,\lambda}^{a_1\ldots a_{s'}}$ with $\lambda \ne s-s'$ become equal to zero, while
the fields $\phi_{\cur,\lambda}^{a_1\ldots a_{s'}}$ with $\lambda=s-s'$ and $s'=0,1,\ldots
s-1$ are expressed in terms of the rank-$s$ traceless tensor field $\phi_{\cur,0}^{a_1\ldots
a_s}$,
\beq
&& \hspace{-0.8cm} \phi_{\cur,\lambda}^{a_1\ldots a_{s'}} = 0 \,,\qquad
\hbox{ for } \quad \lambda \ne s-s'\,,
\\
&& \hspace{-0.8cm} \phi_{\cur,s-s'}^{a_1\ldots a_{s'}} = \sqrt{(s-s')!}
X_\cur
\partial^{b_1} \ldots \partial^{b_{s-s'}} \phi_{\cur,0}^{b_1 \ldots b_{s-s'} a_1
\ldots a_{s'} }\,.
\nonumber\\[-6pt]
\eeq

We recall that, in the standard CFT, the spin-$s$ anomalous conformal current is described by
the traceless field $\phi_{\cur,0}^{a_1\ldots a_s}$. Thus we see that making use of the gauge
symmetry and differential constraint we reduce the field content of our approach to the one
in the standard CFT. In other words, use of the gauge symmetry and differential constraint
allows us to match our approach and the standard formulation of
the spin-$s$ anomalous conformal current%
\footnote{ We note that, as in the standard CFT, our currents can be
considered either as composite operators or as fundamental field degrees of
freedom. At the group theoretical level, we study in this paper, this
distinction does not matter. Methods for building conformal currents as
composite operators are discussed in Refs.\cite{Konstein:2000bi}.}.
To summarize, our gauge invariant approach is equivalent to the standard one.

\subsection{ Light-cone gauge frame }

We now discuss the spin-$s$ anomalous conformal current in the light-cone
gauge frame. To this end we note that, for the anomalous conformal current,
the light-cone gauge frame is achieved through the use of differential
constraint \rf{man01-10112011-09} and light-cone gauge condition. Using gauge
symmetry of the spin-$s$ anomalous conformal current \rf{man01-10112011-22},
we impose the light-cone gauge on the $|\phi_\cur\rangle$,
\be \label{man01-01-12102011-01} \bar\alpha^+ \Pi^\smponetwo
|\phi_\cur\rangle =0 \,, \ee
where $\Pi^\smponetwo$ is given in \rf{man01-13102011-10}. Using gauge
\rf{man01-01-12102011-01} and differential constraint \rf{man01-10112011-09}, we find
\beq \label{man01-01-12102011-02}
&& |\phi_\cur\rangle = \exp\Bigl(
-\frac{\alpha^+}{\partial^+}(\bar\alpha^i\partial^i - \eb_{1\,\cur})\Bigr)
|\phi_\cur^{\rm l.c.}\rangle\,,\qquad
\\
\label{man01-02-17072009-10x3} && \bar\alpha^i \bar\alpha^i | \phi_\cur^{\rm
l.c.}\rangle  = 0 \,,
\eeq
where a light-cone ket-vector $|\phi_\cur^{\rm l.c.}\rangle$ is obtained from
$|\phi_\cur\rangle$ \rf{man01-10112011-05} by equating $\alpha^+ = \alpha^- =
0$,
\be \label{man01-02-17072009-10x4} |\phi_\cur^{\rm l.c.}\rangle \equiv
|\phi_\cur\rangle\Bigr|_{\alpha^+=\alpha^-=0} \,.\ee
We see that we are left with light-cone fields
\be \label{man01-16102011-09}
\phi_{\cur,\lambda}^{i_1\ldots i_{s'}}\,,
\qquad
s'=0,1,\ldots, s\,,
\qquad
\lambda \in  [s-s']_2\,,
\ee
which are traceless tensor fields of $so(d-2)$ algebra, $\phi_{\cur,\lambda}^{iii_3\ldots
i_{s'}}= 0$. These fields constitute the field content of the light-cone gauge frame. Note
that, in contrast to the Stueckelberg gauge frame, all fields $\phi_{\cur,\lambda}^{i_1\ldots
i_{s'}}$ are not equal to zero. Also note that, in contrast to the gauge invariant approach,
the fields \rf{man01-16102011-09} are not subject to any differential constraint.


\section{ Arbitrary spin anomalous shadow field}
\label{sec04}

\subsection{ Gauge invariant formulation }
\label{sec04G}

{\bf Field content}. To discuss gauge invariant formulation of arbitrary
spin-$s$ anomalous shadow field in flat space of dimension $d\geq 4$ we use
the following fields:
\be \label{man01-10112011-01sh}
\phi_{\sh,\lambda}^{a_1\ldots a_{s'}}\,,
\qquad
s'=0,1,\ldots, s\,,
\qquad
\lambda \in  [s-s']_2\,.
\ee
We note that:\\
{\bf i}) In \rf{man01-10112011-01sh}, the fields $\phi_{\sh,\lambda}$ and
$\phi_{\sh,\lambda}^a$ are the respective scalar and vector fields of the Lorentz algebra,
while the field $\phi_{\sh,\, \lambda}^{a_1\ldots a_{s'}}$, $s'>1$, is rank-$s'$ totally
symmetric traceful tensor field of the Lorentz algebra $so(d-1,1)$. Using shortcut
$\phi_\lambda^{s'}$ for the field $\phi_{\sh,\lambda}^{a_1\ldots a_{s'}}$, fields in
\rf{man01-10112011-01sh} can be represented as in \rf{man01-16102011-08}.
\\
{\bf ii}) The tensor fields $\phi_{\sh,\lambda}^{a_1\ldots a_{s'}}$ with
$s'\geq 4 $ satisfy the double-tracelessness constraint
\be \label{man01-10112011-02sh} \phi_{\sh,\lambda}^{aabba_5\ldots
a_{s'}}=0\,, \qquad s'=4,5,\ldots, s\,. \ee
\\
{\bf iii}) The fields $\phi_{\sh,\lambda}^{a_1\ldots a_{s'}}$ have the
following conformal dimensions:
\be \label{man01-10112011-03sh}
\Delta(\phi_{\sh,\lambda}^{a_1\ldots a_{s'}})= \frac{d}{2} - \kappa +
\lambda\,, \ee
{\bf iv)} In the framework of AdS/CFT correspondence, $\kappa$ is related to the mass
parameter $m$ of spin-$s$ massive field in $AdS_{d+1}$ as in \rf{man01-11102011-04}.

In order to obtain the gauge invariant description in an easy--to--use form
we use the oscillators and introduce a ket-vector $|\phi_\sh\rangle$ defined
by
\beq \label{man01-10112011-05sh}
&& |\phi_\sh\rangle = \sum_{s'=0}^s |\phi_\sh^{s'}\rangle \,,
\\
&& |\phi_\sh^{s'}\rangle
\nonumber\\
&&  = \!\!\! \sum_{\lambda \in [s-s']_2}\!\!
\frac{\zeta_{\phantom{z}}^{\frac{s-s'-\lambda}{2}}
\alpha_z^{\frac{s-s'+\lambda}{2}}\alpha^{a_1}\ldots
\alpha^{a_{s'}}}{s'!\sqrt{(\frac{s-s'+\lambda}{2})!
(\frac{s-s'-\lambda}{2})!}} \, \phi_{\sh,\lambda}^{a_1\ldots a_{s'}}
|0\rangle\,.
\nonumber
\eeq
From \rf{man01-10112011-05sh}, we see that the ket-vector $|\phi_\sh\rangle$
is degree-$s$ homogeneous polynomial in the oscillators $\alpha^a$,
$\alpha^z$, $\zeta$, while the ket-vector $|\phi_\sh^{s'}\rangle$ is
degree-$s'$ homogeneous polynomial in the oscillators $\alpha^a$, i.e., these
ket-vectors satisfy the relations
\beq \label{man01-10112011-06sh}
&& (N_\alpha + N_z + N_\zeta -s) |\phi_\sh\rangle =0 \,,
\\
\label{man01-10112011-07sh} && (N_\alpha -s') |\phi_\sh^{s'}\rangle =0\,.
\eeq
In terms of the ket-vector $|\phi_\sh\rangle$, double-tracelessness
constraint \rf{man01-10112011-02sh} takes the form
\beq \label{man01-10112011-08sh}
&& (\bar{\alpha}^2)^2 |\phi_\sh\rangle  = 0 \,.
\eeq

{\bf Differential constraint}. We find the following differential constraint
for the anomalous shadow field:
\beq \label{man01-10112011-09sh}
&& \Cb_\sh |\phi_\sh\rangle = 0 \,,
\\
\label{man01-10112011-10sh} && \Cb_\sh  = \albpar - \half \alpar \bar\alpha^2 -
\eb_{1,\sh} \Pi^\smponetwo + \half e_{1,\sh} \bar\alpha^2 \,,\qquad\quad
\\
\label{man01-10112011-12sh} && e_{1,\sh} =  \zeta r_\zeta  + \alpha^zr_z \Box
\,,
\nonumber\\[-10pt]
&&
\\[-10pt]
&& \eb_{1,\sh} = -r_\zeta\bar\zeta \Box  - r_z \bar\alpha^z \,,
\nonumber
\eeq
where the operators $\Pi^\smponetwo$, $r_\zeta$, $r_z$ are defined in
\rf{man01-13102011-10}, \rf{man01-13102011-11}. One can make sure that
constraint \rf{man01-10112011-09sh} is invariant under gauge transformation
and conformal algebra transformations which we discuss below.

{\bf Gauge symmetries}. We now discuss gauge symmetries of the anomalous
shadow field. To this end we introduce the following gauge transformation
parameters:
\be \label{man01-10112011-15sh}
\xi_{\sh,\lambda}^{a_1\ldots a_{s'}}\,,
\qquad
s'=0,1,\ldots, s-1\,,
\qquad
\lambda \in [s-1-s']_2\,.
\ee
We note that\\
{\bf i}) In \rf{man01-10112011-15sh}, the gauge transformation parameters $\xi_{\sh,\lambda}$
and $\xi_{\sh,\lambda}^a$ are the respective scalar and vector fields of the Lorentz algebra,
while the gauge transformation parameter $\xi_{\sh,\lambda}^{a_1\ldots a_{s'}}$, $s'>1$, is
rank-$s'$ totally symmetric tensor field of the Lorentz algebra $so(d-1,1)$.
\\
{\bf ii}) The gauge transformation parameters $\xi_{\sh,\lambda}^{a_1\ldots
a_{s'}}$ with $s'\geq 2 $ satisfy the tracelessness constraint
\be \label{man01-10112011-16sh} \xi_{\sh,\lambda}^{aaa_3\ldots a_{s'}}=0\,,
\qquad s'= 2,3,\ldots, s-1\,. \ee
{\bf iii}) The gauge transformation parameters $\xi_{\sh,\lambda}^{a_1\ldots
a_{s'}}$ have the conformal dimensions
\be \label{man01-10112011-17sh}
\Delta(\xi_{\sh,\lambda}^{a_1\ldots a_{s'}})= \frac{d}{2} - \kappa + \lambda
-1\,.
\ee

Now, as usually, we collect the gauge transformation parameters in a ket-vector
$|\xi_\sh\rangle$ defined by
\beq \label{man01-10112011-18sh}
&& |\xi_\sh\rangle  = \sum_{s'=0}^{s-1} |\xi_\sh^{s'}\rangle\,,
\\
&& |\xi_\sh^{s'}\rangle
\nonumber\\
&& =\!\!\!\!\sum_{\lambda \in [s-1-s']_2}\!\!\!
\frac{\zeta_{\phantom{z}}^{\frac{s-1-s'-\lambda}{2}}
\alpha_z^{\frac{s-1-s'+\lambda}{2}}\alpha^{a_1}\ldots
\alpha^{a_{s'}}}{s'!\sqrt{(\frac{s-1-s'+\lambda}{2})!
(\frac{s-1-s'-\lambda}{2})!}} \, \xi_{\sh,\lambda}^{a_1\ldots a_{s'}}
|0\rangle\,.
\nonumber
\eeq
The ket-vectors $|\xi_\sh\rangle$, $|\xi_\sh^{s'}\rangle$ satisfy the
algebraic constraints
\beq \label{man01-10112011-19sh}
&& (N_\alpha + N_z + N_\zeta - s +1 ) |\xi_\sh\rangle = 0 \,,
\\
\label{man01-10112011-20sh} && (N_\alpha -s')|\xi_\sh^{s'}\rangle  = 0\,,
\eeq
which tell us that $|\xi_\sh\rangle$ is a degree-$(s-1)$ homogeneous
polynomial in the oscillators $\alpha^a$, $\alpha^z$, $\zeta$, while the
ket-vector $|\xi_\sh^{s'}\rangle$ is degree-$s'$ homogeneous polynomial in
the oscillators $\alpha^a$. In terms of the ket-vector $|\xi_\sh\rangle$,
tracelessness constraint \rf{man01-10112011-16sh} takes the form
\be \label{man01-10112011-21sh} \bar\alpha^2 |\xi_\sh\rangle = 0 \,.\ee

Gauge transformation can entirely be written in terms of $|\phi_\sh\rangle$
and $|\xi_\sh\rangle$. This is to say that gauge transformation takes the
form
\beq
\label{man01-10112011-22sh} && \delta |\phi_\sh\rangle = G_\sh
|\xi_\sh\rangle\,,
\\
&& G_\sh  =\alpar - e_{1,\sh} - \alpha^2 \frac{1}{2N_\alpha + d- 2}
\eb_{1,\sh} \,,\qquad
\eeq
where $e_{1,\sh}$, $\eb_{1,\sh}$ are given in \rf{man01-10112011-12sh}.
Constraint \rf{man01-10112011-09sh} is invariant under gauge transformation
\rf{man01-10112011-22sh}.

{\bf Realization of conformal algebra symmetries}. To complete the gauge invariant
formulation of the spin-$s$ anomalous shadow field we provide realization of the conformal
algebra symmetries on space of the ket-vector $|\phi_\sh\rangle$. All that is required is to
fix the operators $M^{ab}$, $\Delta$, and $R^a$ and insert then these operators into
\rf{conalggenlis01}-\rf{conalggenlis04}. Realization of the spin operator $M^{ab}$ on
ket-vector $|\phi_\sh\rangle$ \rf{man01-10112011-05sh} is given in \rf{mabdef0001}, while
realization of the operator $\Delta$,
\be \label{man01-15102011-59x1}
\Delta_\sh = \frac{d}{2} -\nu\,,
\qquad
\nu = \kappa + N_\zeta - N_z \,,
\ee
can be read from \rf{man01-10112011-03sh}. Realization of the operator $R^a$ on space of
$|\phi_\sh\rangle$, which we find, is given by
\beq \label{man01-15102011-61}
R_\sh^a  & = & 2 \alpha^z r_z \Bigl( (\nu-1) \bar\alpha^a  +
\bar{C}_\perp^a\Bigr)
\nonumber\\
& + & 2\Bigl( \nu \Cwt^a - \alpha^2 \frac{1}{2N_\alpha+d-2}
\Cb_\perp^a \Bigr)r_\zeta\bar\zeta\,,
\eeq
where the operators  $\Cwt^a$, $\Cb_\perp^a$ are given in
\rf{man01-16102011-05},\rf{man01-16102011-06}, while the operators $r_\zeta$, $r_z$ are
defined in \rf{man01-13102011-11}.

{\bf Two-point gauge invariant vertex}. We now discuss two-point vertex for
the spin-$s$ anomalous shadow field. This is to say that we find the
following gauge invariant two-point vertex:
\beq \label{10112011-03}
&& \Gamma = \int d^dx_1 d^dx_2 \Gamma_{12}\,,
\\
\label{man01-14102011-02} && \Gamma_{12}  = \half \langle\phi_\sh(x_1)|
\frac{\mubf f_\nu}{ |x_{12}|^{2\nu + d }} |\phi_\sh (x_2)\rangle \,,
\\
\label{man01-15102011-11}
&& f_\nu = \frac{\Gamma(\nu + \frac{d}{2})\Gamma(\nu + 1)}{4^{\kappa - \nu}
\Gamma(\kappa + \frac{d}{2})\Gamma(\kappa + 1)} \,,
\\
&&  \nu = \kappa + N_\zeta -N_z\,,
\\
\label{manus2009-02-03}
&& |x_{12}|^2 \equiv x_{12}^a x_{12}^a\,, \qquad x_{12}^a = x_1^a - x_2^a\,,
\eeq
where $\mubf$ is given in \rf{man01-16102011-04}. Vertex $\Gamma$ \rf{10112011-03} is
invariant under gauge transformation of the anomalous shadow field $|\phi_\sh\rangle$
\rf{man01-10112011-22sh} provided this anomalous shadow field satisfies differential
constraint \rf{man01-10112011-09sh}. The vertex is obviously invariant under the Poincar\'e
algebra and dilatation symmetries. We make sure that vertex \rf{10112011-03} is invariant
under the conformal boost transformations.

To illustrate structure of the vertex $\Gamma_{12}$ we note that, in terms of
the tensor fields $\phi_{\sh,\lambda}^{a_1\ldots a_{s'}}$, vertex
$\Gamma_{12}$ \rf{man01-14102011-02} can be represented as
\beq
&&  \hspace{-1cm} \Gamma_{12} = \sum_{s'=0}^s
\sum_{\lambda \in [s-s']_2}\Gamma_{12,\,\lambda}^{s'}\,,
\\[5pt]
&&  \hspace{-1cm} \Gamma_{12,\lambda}^{s'} =
\frac{w_\lambda}{2s'!|x_{12}|^{2\kappa -2\lambda +d }}\Bigr(
\phi_{\sh,\lambda}^{a_1\ldots a_{s'}} (x_1)\phi_{\sh,\lambda}^{a_1\ldots
a_{s'}}(x_2)
\nonumber\\[5pt]
&&  \hspace{-0.2cm} - \frac{s'(s'-1)}{4} \phi_{\sh,\lambda}^{aaa_3\ldots
a_{s'}}(x_1) \phi_{\sh,\lambda}^{bba_3\ldots a_{s'}}(x_2)\Bigl)\,,
\\[10pt]
\label{man01-15102011-14} &&  \hspace{-1cm} w_\lambda \equiv
\frac{\Gamma(\kappa -\lambda + \frac{d}{2})\Gamma(\kappa -\lambda
+1)}{4^\lambda \Gamma(\kappa +\frac{d}{2})\Gamma(\kappa+1)} \,.
\eeq

We note that the kernel of the vertex $\Gamma$ is connected with a two-point correlation
function of the anomalous conformal current. In the framework of our approach, the anomalous
conformal current is described by gauge fields \rf{man01-10112011-01} subject to differential
constraints. In order to discuss the correlation function of the spin-$s$ anomalous conformal
current in a proper way, we can impose a gauge condition on the gauge fields given in
\rf{man01-10112011-01}. Recall that we have considered the spin-$s$ anomalous conformal
current by using the Stueckelberg and light-cone gauge frames. Obviously, the two-point
correlation function of the anomalous conformal current in the Stueckelberg and light-cone
gauge frames is obtained from the two-point vertex $\Gamma$ taken in the respective
Stueckelberg and light-cone gauge frames. To this end we proceed by discussing the anomalous
shadow field in the Stueckelberg and light-cone gauge frames.

\subsection{ Stueckelberg gauge frame }
\label{sec04S}

We now discuss the spin-$s$ anomalous shadow field in the Stueckelberg gauge frame. We note
that  the Stueckelberg gauge frame can be achieved through the use of differential constraint
given in \rf{man01-10112011-09sh} and the Stueckelberg gauge condition. From
\rf{man01-10112011-22sh}, we see that a field defined by
\be \bar\zeta \Pi^\smponetwo |\phi_\sh\rangle \ee
transforms as Stueckelberg field. Therefore this field can be gauged away via Stueckelberg
gauge fixing,
\be \label{man01-14102011-01} \bar\zeta \Pi^\smponetwo |\phi_\sh\rangle = 0
\,. \ee
Using this gauge condition and differential constraint
\rf{man01-10112011-09sh}, we find the following relations:
\beq  \label{10112011-04sh}
&& \hspace{-0.8cm}  \bar\alpha^2 |\phi_\sh^{s'}\rangle =0\,,
\\
\label{10112011-05sh} &&  \hspace{-0.8cm} |\phi_\sh^{s'}\rangle  = X_\sh
\alpha_z^{s-s'} (\albpar)^{s-s'} |\phi_\sh^s\rangle\,,
\\
&&  \hspace{-0.8cm} X_\sh \!\equiv\! \frac{(-)^{s-s'}}{(s-s')!} \Bigl(
\frac{2^{s-s'} \Gamma(\kappa\!+\! \frac{d-2}{2}\!+\!s')
\Gamma(\kappa\!+\!1)}{\Gamma(\kappa\!+\!s+\!\frac{d-2}{2})
\Gamma(\kappa\!-\!s\!+\!1+\!s')}\!\Bigr)^{1/2}\!\!.
\nonumber
\eeq
Relation \rf{10112011-04sh} tells us that all fields $\phi_{\sh,\lambda}^{a_1\ldots a_{s'}}$
with $s'\geq 2$ become traceless. From relation \rf{10112011-05sh}, we learn that the fields
$\phi_{\sh,\lambda}^{a_1\ldots a_{s'}}$ with $\lambda \ne s-s'$ become equal to zero, while
the fields $\phi_{\sh,\lambda}^{a_1\ldots a_{s'}}$ with $\lambda=s-s'$ and $s'=0,1,\ldots,
s-1$ are expressed in terms of the rank-$s$ traceless tensor field $\phi_{\sh,0}^{a_1\ldots
a_s}$,
\beq
&& \hspace{-0.8cm} \phi_{\sh,\lambda}^{a_1\ldots a_{s'}} = 0 \,,\qquad \hbox{
for } \quad \lambda \ne s-s'\,,
\\
&& \hspace{-0.8cm} \phi_{\sh,s-s'}^{a_1\ldots a_{s'}} = \sqrt{(s-s')!} X_\sh
\partial^{b_1} \ldots \partial^{b_{s-s'}} \phi_{\sh,0}^{b_1 \ldots b_{s-s'} a_1
\ldots a_{s'} }\,.
\nonumber\\[-6pt]
\eeq

We recall that, in the standard CFT, the spin-$s$ anomalous shadow field is described by the
traceless rank-$s$ tensor field $\phi_{\sh,0}^{a_1\ldots a_s}$. Thus we see that making use
of the gauge symmetry and differential constraint we reduce the field content of our approach
to the one in the standard CFT. In other words, use of the gauge symmetry and differential
constraint allows us to match our approach and the standard formulation of the anomalous
shadow field. To summarize, our gauge invariant approach is equivalent to the standard one.

We now discuss Stueckelberg gauge-fixed two-point vertex of the anomalous shadow field. In
other words, we are going to connect our vertex \rf{10112011-03} with the one in the standard
CFT. To do that we note that vertex of the standard CFT is obtained from our gauge invariant
vertex \rf{10112011-03} by plugging solution to the differential constraint
\rf{10112011-05sh} into \rf{10112011-03}. Doing so, we find the following two-point density
(up to total derivative) in the Stueckelberg gauge frame:
\beq \label{man01-14102011-03}
&& \hspace{-1cm}\Gamma_{12}^{{\rm Stuck.g.fram}} = k_s \Gamma_{12}^{{\rm
stand}}\,,
\\[9pt]
\label{man01-14102011-04} && \hspace{-0.5cm} \Gamma_{12}^{{\rm stand}} = s!
\langle \phi_\sh^s(x_1)| \Obf_{12}|\phi_\sh^s(x_2)\rangle \,,\quad
\\[9pt]
&& \hspace{-0.5cm} \Obf_{12} \equiv \sum_{n=0}^s
\frac{(-)^n 2^n}{n!} \frac{(\alpha x_{12})^n (\bar\alpha
x_{12})^n}{|x_{12}|^{2\kappa +d + 2n}}\,,
\\[7pt]
\label{man01-15102011-54} && \hspace{-0.5cm} k_s \equiv
\frac{2\kappa+2s+d-2}{2s!(2\kappa+d-2)}\,,
\eeq
where $\alpha x_{12} = \alpha^a x_{12}^a$, $\bar\alpha x_{12} = \bar\alpha^a x_{12}^a$ and
$\Gamma_{12}^{{\rm stand}}$ in \rf{man01-14102011-03}, \rf{man01-14102011-04} stands for the
two-point vertex of the spin-$s$ anomalous shadow field in the standard CFT. Relation
\rf{man01-14102011-04} provides oscillator representation for the $\Gamma_{12}^{{\rm
stand}}$. In terms of the tensor field $\phi_{\sh,0}^{a_1 \ldots a_s}$, vertex
$\Gamma_{12}^{{\rm stand}}$ \rf{man01-14102011-04} can be represented in the commonly used
form,
\beq \label{man01-15102011-53}
&& \hspace{-0.8cm} \Gamma_{12}^{{\rm stand}} = \phi_{\sh,0}^{a_1 \ldots a_s
}(x_1) \frac{O_{12}^{a_1b_1} \ldots O_{12}^{a_s b_s}}{|x_{12}|^{2\kappa+d}}
\phi_{\sh,0}^{b_1\ldots b_s}(x_2) \,,\qquad
\\
\label{man01-14102011-05} && \hspace{0.3cm}  O_{12}^{ab}  \equiv \eta^{ab} -
\frac{2x_{12}^a x_{12}^b}{|x_{12}|^2}\,.
\eeq
From \rf{man01-14102011-03}, we see that our gauge invariant vertex
$\Gamma_{12}$ considered in the Stueckelberg gauge frame coincides, up to
normalization factor $k_s$, with the two-point vertex in the standard CFT. In
section \ref{sec03S}, we have demonstrated that, in the Stueckelberg gauge
frame, we are left with rank-$s$ traceless tensor field
$\phi_{\cur,0}^{a_1\ldots a_s}$. Two-point correlation function of this
tensor field is defined by the kernel of vertex $\Gamma^\stand$
\rf{man01-15102011-53}.

\subsection{ Light-cone gauge frame }

We proceed with discussion of the anomalous shadow field in the light-cone
gauge frame. For the anomalous shadow field, the light-cone gauge frame can
be achieved through the use of differential constraint
\rf{man01-10112011-09sh} and light-cone gauge condition. Using gauge symmetry
of the spin-$s$ anomalous shadow field \rf{man01-10112011-22sh}, we impose
the light-cone gauge on the $|\phi_\sh\rangle$,
\be \label{man01-01-12102011-01sh} \bar\alpha^+ \Pi^\smponetwo |\phi_\sh\rangle
=0 \,, \ee
where $\Pi^\smponetwo$ is given in \rf{man01-13102011-10}. Using gauge
condition \rf{man01-01-12102011-01sh} and differential constraint
\rf{man01-10112011-09sh}, we obtain
\beq \label{man01-01-12102011-02sh}
&& |\phi_\sh\rangle = \exp\Bigl(
-\frac{\alpha^+}{\partial^+}(\bar\alpha^i\partial^i - \eb_{1\,\sh})\Bigr)
|\phi_\sh^{\rm l.c.}\rangle\,,\qquad
\\[5pt]
\label{man01-02-17072009-10x3sh} && \bar\alpha^i \bar\alpha^i | \phi_\sh^{\rm
l.c.}\rangle  = 0 \,,
\eeq
where a light-cone ket-vector $|\phi_\sh^{\rm l.c.}\rangle$ is obtained from
$|\phi_\sh\rangle$ \rf{man01-10112011-05sh} by equating $\alpha^+ = \alpha^- =
0$,
\be \label{man01-02-17072009-10x4sh} |\phi_\sh^{\rm l.c.}\rangle \equiv
|\phi_\sh\rangle\Bigr|_{\alpha^+=\alpha^-=0} \,.\ee
We see that we are left with light-cone fields
\be \label{man01-16102011-10}
\phi_{\sh,\lambda}^{i_1\ldots i_{s'}}\,,
\qquad
s'=0,1,\ldots, s\,,
\qquad
\lambda \in [s-s']_2\,,
\ee
which are traceless tensor fields of $so(d-2)$ algebra, $\phi_{\sh,\lambda}^{iii_3\ldots
i_{s'}}= 0$. These fields constitute the field content of the light-cone gauge frame. Note
that, in contrast to the Stueckelberg gauge frame, all fields $\phi_{\sh,\lambda}^{i_1\ldots
i_{s'}}$ are not equal to zero. Also note that, in contrast to the gauge invariant approach,
fields \rf{man01-16102011-10} are not subject to any differential constraint. Using
\rf{man01-01-12102011-02sh} in \rf{man01-14102011-02} leads to light-cone gauge-fixed vertex
\beq \label{man01-15102011-10}
\Gamma_{12}^{\rm l.c.} &  = &  \half \langle\phi_\sh^{\rm l.c.}(x_1)|
\frac{f_\nu}{ |x_{12}|^{2\nu + d }} |\phi_\sh^{\rm l.c.}(x_2)\rangle \,,
\qquad
\eeq
where $f_\nu$ is defined in \rf{man01-15102011-11}.

To illustrate the structure of vertex $\Gamma_{12}^{\rm l.c.}$
\rf{man01-15102011-10} we note that, in terms of the fields
$\phi_{\sh,\lambda}^{i_1\ldots i_{s'}}$, the vertex can be represented as
\beq  \label{man01-15102011-15}
&&  \hspace{-1cm} \Gamma_{12}^{\rm l.c.} = \sum_{s'=0}^s \sum_{\lambda \in
[s-s']_2}\Gamma_{12,\lambda}^{s'\, {\rm l.c.}}\,,
\\[5pt]
&&  \hspace{-1cm} \Gamma_{12,\lambda}^{s'\,{\rm l.c.}}\!=\!
\frac{w_\lambda}{2s'!|x_{12}|^{2\kappa -2\lambda +d }}
\phi_{\sh,\lambda}^{i_1\ldots
i_{s'}}\!(\!x_1\!)\!\phi_{\sh,\lambda}^{i_1\ldots i_{s'}}\!(\!x_2\!),
\eeq
where $w_\lambda$ is given in \rf{man01-15102011-14}. We see that, as in the
case of gauge invariant vertex, light-cone vertex \rf{man01-15102011-15} is
diagonal with respect to the light-cone fields $\phi_{\sh,\lambda}^{i_1\ldots
i_{s'}}$. Note however that, in contrast to the gauge invariant vertex, the
light-cone vertex is constructed out of the light-cone fields which are not
subject to any differential constraints.

Thus, we see that our gauge invariant vertex does indeed provide easy and
quick access to the light-cone gauge vertex. Namely, all that is needed to
obtain light-cone gauge vertex \rf{man01-15102011-15} is to remove traces of
the tensor fields $\phi_{\sh,\lambda}^{a_1\ldots a_{s'}}$ and replace the
$so(d-1,1)$ Lorentz algebra vector indices appearing in gauge invariant
vertex \rf{man01-14102011-02} by the respective vector indices of the
$so(d-2)$ algebra.

The kernel of the light-cone vertex gives the two-point correlation function of the spin-$s$
anomalous conformal current taken to be in the light-cone gauge. Defining two-point
correlation functions of the fields $\phi_{\cur,\lambda}^{i_1\ldots i_{s'}}$ as the second
functional derivative of $\Gamma$ with respect to the shadow fields
$\phi_{\sh,-\lambda}^{i_1\ldots i_{s'}}$, we obtain the following correlation functions:
\beq
&& \langle \phi_{\cur,\lambda}^{i_1\ldots i_{s'}}(x_1), \phi_{\cur,\lambda}^{j_1\ldots
j_{s'}}(x_2)\rangle
\nonumber\\[5pt]
&&\qquad\qquad  = \frac{w_{-\lambda}}{|x_{12}|^{2\kappa + 2\lambda +d}} \Pi^{i_1\ldots
i_{s'};j_1\ldots j_{s'}}\,, \qquad
\eeq
where $w_\lambda$ is defined in \rf{man01-15102011-14}  and $\Pi^{i_1\ldots i_{s'};j_1\ldots
j_{s'}}$ stands for the projector on traceless rank-$s'$ tensor field of the $so(d-2)$
algebra. Explicit form of the projector may be found, e.g., in Ref.\cite{Erdmenger:1997wy}.

\section{ Two-point current-shadow field interaction
vertex}
\label{sec09}

We now briefly discuss the two-point current-shadow field interaction vertex. In our
approach, this interaction vertex is determined by requiring that:
\\
{\bf i}) the vertex is invariant under both gauge transformations of anomalous conformal
current and shadow field;
\\
{\bf ii}) vertex is invariant under conformal algebra transformations.

We find the following vertex:
\be  \label{man01-15102011-16}   \LL = \langle \phi_\cur|\mubf |
\phi_\sh\rangle \,, \ee
where $\mubf$ is given in \rf{man01-16102011-04}. We note that, under gauge
transformation of the anomalous conformal current \rf{man01-10112011-22}, the
variation of vertex \rf{man01-15102011-16} takes the form (up to total
derivative)
\be \label{oldman-23012012-01} \delta_{ \xi_\cur} \LL = - \langle
\xi_\cur|\Cb_\sh|\phi_\sh\rangle\,. \ee
From \rf{oldman-23012012-01}, we see that the vertex $\LL$ is invariant under gauge
transformation of the anomalous conformal current provided the anomalous shadow field
satisfies differential constraint \rf{man01-10112011-09sh}. Next, we note that under gauge
transformation of the anomalous shadow field \rf{man01-10112011-22sh} the gauge variation of
vertex \rf{man01-15102011-16} takes the form (up to total derivative)
\be  \label{oldman-23012012-02}   \delta_{\xi_\sh}\LL =  -\langle\xi_\sh|
\Cb_\cur|\phi_\cur\rangle \,. \ee
From \rf{oldman-23012012-02}, we see that the vertex $\LL$ is invariant under
gauge transformation of the anomalous shadow field provided the anomalous
conformal current satisfies differential constraint \rf{man01-10112011-09}.

Using the realization of the conformal algebra symmetries obtained in
Sections \ref{sec03},\ref{sec04}, we make sure that vertex $\LL$
\rf{man01-15102011-16} is invariant under the conformal algebra
transformations.

\section{AdS/CFT correspondence. Preliminaries}\label{secAdS/CFT}

We now study the AdS/CFT correspondence for free arbitrary spin massive AdS field and
boundary arbitrary spin anomalous conformal current and shadow field. To study the AdS/CFT
correspondence we use the gauge invariant CFT adapted formulation of massive AdS field and
modified de Donder gauge
condition found in Ref.\cite{Metsaev:2009hp}.%
\footnote{ Applications of the standard de Donder gauge to the various problems of massless
fields may be found in Refs.\cite{Guttenberg:2008qe}. Recent interesting discussion of
modified de Donder gauge may be found in Ref.\cite{Chang:2011mz}. We believe that our
modified de Donder gauge will also be useful for better understanding of various aspects of
AdS/QCD correspondence which are discussed, e.g., in Refs.\cite{Brodsky:2008pg}.}
We emphasize that it is the use of our massive gauge fields and the modified de Donder gauge
condition that leads to the decoupled gauge-fixed equations of motion
and surprisingly simple Lagrangian%
\footnote{ Our massive gauge fields are obtained from gauge fields used in gauge invariant
approach to massive fields in Ref.\cite{Zinoviev:2001dt} by the invertible transformation
which is described in Appendix in Ref.\cite{Metsaev:2009hp}. Discussion of interesting
methods for solving AdS field equations of motion without gauge fixing may be found in
Refs.\cite{Bolotin:1999fa}.}. The use of our massive gauge fields and the modified de Donder
gauge condition makes the study of AdS/CFT correspondence for arbitrary spin-$s$ massive AdS
field similar to the one for spin-0 massive AdS field. Owing these properties of our massive
gauge fields and the modified de Donder gauge condition, the computation of effective action
is considerably simplified. Perhaps, this is the main advantage of our approach.

In our approach to the AdS/CFT correspondence, we have gauge symmetries not
only at AdS side but also at the boundary CFT. Also, we note that the
modified de Donder gauge condition turns out to be invariant under on-shell
leftover gauge symmetries of massive AdS field. This is to say that, in the
framework of our approach, the study of AdS/CFT correspondence implies the
matching of:
\\
{\bf i}) modified de Donder gauge condition for bulk massive field and
the corresponding differential constraint for boundary anomalous conformal current and shadow
field;
\\
{\bf ii}) on-shell leftover gauge symmetries of bulk massive field and
the corresponding gauge symmetries of boundary anomalous conformal current and shadow field;
\\
{\bf iii}) on-shell global symmetries of bulk massive field and the corresponding global
symmetries of boundary anomalous conformal current and shadow field;
\\
{\bf iv}) an effective action evaluated on the solution of AdS massive field equations of
motion with the Dirichlet problem corresponding to the boundary anomalous
shadow field and the boundary two-point gauge invariant vertex for the anomalous
shadow field.

As we have already said, to discuss the AdS/CFT correspondence for bulk arbitrary spin
massive AdS field and boundary arbitrary spin anomalous conformal current and shadow field we
use the CFT adapted gauge invariant Lagrangian and the modified de Donder gauge condition for
the arbitrary spin massive AdS field found in Ref.\cite{Metsaev:2009hp}. We begin therefore
with the presentation of our result in Ref.\cite{Metsaev:2009hp}.

\subsection{ CFT adapted approach to massive arbitrary spin AdS field
}\label{man02-sec-06}

In $AdS_{d+1}$ space, massive spin-$s$ field is described by the following
scalar, vector, and totally symmetric tensor fields of the $so(d)$ algebra:%
\footnote{ From now on we use, unless otherwise specified, the Euclidian
signature.}
\be \label{man01-15102011-19}
\phi_\lambda^{a_1\ldots a_{s'}}\,,
\qquad
s'=0,1,\ldots, s\,,
\qquad
\lambda \in [s-s']_2\,.
\ee
Using shortcut $\phi_\lambda^{s'}$ for the field $\phi_\lambda^{a_1\ldots
a_{s'}}$, fields in \rf{man01-15102011-19} can be represented as in
\rf{man01-16102011-08}. The fields $\phi_\lambda^{a_1\ldots a_{s'}}$ with $s'
\geq 4$ are double-traceless,
\be \label{man01-15102011-20} \phi_\lambda^{aabba_5\ldots a_{s'}}=0\,,
\hspace{1cm} s'=4,5,\ldots,s. \ee

In order to obtain the gauge invariant description in an easy--to--use form
we use the oscillators and introduce a ket-vector $|\phi\rangle$ defined by
\beq \label{man01-15102011-21}
&& |\phi\rangle = \sum_{s'=0}^s |\phi^{s'}\rangle \,,
\\
&& |\phi^{s'}\rangle
\nonumber\\
&&  = \!\!\! \sum_{\lambda \in [s-s']_2}\!\!
\frac{\zeta_{\phantom{z}}^{\frac{s-s'+\lambda}{2}}
\alpha_z^{\frac{s-s'-\lambda}{2}}\alpha^{a_1}\ldots
\alpha^{a_{s'}}}{s'!\sqrt{(\frac{s-s'+\lambda}{2})!
(\frac{s-s'-\lambda}{2})!}} \, \phi_\lambda^{a_1\ldots a_{s'}} |0\rangle\,.
\nonumber
\eeq
From \rf{man01-15102011-21}, we see that the ket-vector $|\phi\rangle$ is
degree-$s$ homogeneous polynomial in the oscillators $\alpha^a$, $\alpha^z$,
$\zeta$, while the ket-vector $|\phi^{s'}\rangle$ is degree-$s'$ homogeneous
polynomial in the oscillators $\alpha^a$. In terms of the ket-vector
$|\phi\rangle$, double-tracelessness constraint \rf{man01-15102011-20} takes
the form
\beq
\label{man01-15102011-22} && (\bar{\alpha}^2)^2 |\phi\rangle  = 0 \,.
\eeq

Using the Poinca\'e parametrization of $AdS_{d+1}$ space
\be \label{lineelem01} ds^2 = \frac{1}{z^2}(dx^a dx^a + dz\, dz)\,, \ee
we present CFT adapted gauge invariant action and Lagrangian
\cite{Metsaev:2009hp},
\beq \label{man01-15102011-23a1}
S & = & \int d^dx dz \LL \,,
\\
\label{man01-15102011-23}
\LL &= &  \half \langle \partial^a \phi|\mubf | \partial^a \phi\rangle +\half
\langle \TT_{\nu-\half} \phi| \mubf | \TT_{\nu-\half} \phi\rangle
\nonumber\\
&  - & \half \langle \Cb\phi|| \Cb\phi\rangle\,,
\eeq
where we use the notation
\beq
\label{080405-01add} && \hspace{-0.7cm} \Cb = \albpar - \half \alpar
\bar\alpha^2 - \eb_1\Pi^\smponetwo  + \half e_1 \bar\alpha^2 \,,
\\
\label{e1def01} && - e_1 = \zeta r_\zeta \TT_{ -\nu - \half} + \alpha^z r_z
\TT_{\nu-\half} \,,
\\
&& - \eb_1 = \TT_{\nu + \half}  r_\zeta \bar\zeta  + \TT_{-\nu + \half} r_z
\bar\alpha^z \,,
\\
\label{03072009-01} && \TT_\nu = \partial_z+ \frac{\nu}{z}\,,
\\
\label{man01-16102011-03} && \nu = \kappa + N_\zeta - N_z\,,
\eeq
and $\Pi^\smponetwo$, $\mubf$, $r_\zeta$, $r_z$, and $\kappa$ are given in
\rf{man01-13102011-10}, \rf{man01-16102011-04},  \rf{man01-13102011-11}, and
\rf{man01-11102011-04} respectively.

To discuss gauge symmetries of Lagrangian \rf{man01-15102011-23} we introduce
the gauge transformations parameters,
\be \label{man01-15102011-24}
\xi_\lambda^{a_1\ldots a_{s'}}\,,
\qquad
s'=0,1,\ldots, s-1\,,
\qquad
\lambda \in [s-1-s']_2\,,
\ee
which are scalar, vector, and totally symmetric tensor fields of the $so(d)$
algebra. The gauge transformation parameters with $s'\geq 2$ are traceless,
$\xi_\lambda^{aaa_3\ldots a_{s'}} = 0$.

As usually, we collect the gauge transformation parameters in a ket-vector
$|\xi\rangle$ defined by
\beq \label{man01-15102011-25}
&& |\xi\rangle  = \sum_{s'=0}^{s-1} |\xi^{s'}\rangle\,,
\\
&& |\xi^{s'}\rangle
\nonumber\\
&& =\!\!\!\!\sum_{\lambda \in [s-1-s']_2}\!\!\!
\frac{\zeta_{\phantom{z}}^{\frac{s-1-s'+\lambda}{2}}
\alpha_z^{\frac{s-1-s'-\lambda}{2}}\alpha^{a_1}\ldots
\alpha^{a_{s'}}}{s'!\sqrt{(\frac{s-1-s'+\lambda}{2})!
(\frac{s-1-s'-\lambda}{2})!}} \, \xi_\lambda^{a_1\ldots a_{s'}} |0\rangle\,.
\nonumber
\eeq
We see that $|\xi\rangle$ is a degree-$(s-1)$ homogeneous polynomial in the
oscillators $\alpha^a$, $\alpha^z$, $\zeta$, while the ket-vector
$|\xi^{s'}\rangle$ is degree-$s'$ homogeneous polynomial in the oscillators
$\alpha^a$.

Lagrangian \rf{man01-15102011-23} is invariant under the gauge transformation (up to total derivative)
\beq \label{man01-15102011-26}
&& \delta \phik = G \xik \,,
\\
\label{man01-15102011-26a1} && G \equiv    \alpar - e_1  - \alpha^2
\frac{1}{2N_\alpha+d-2}\eb_1\,.
\eeq

{\bf Global AdS symmetries in CFT adapted approach}. As is well known
relativistic symmetries of fields in $AdS_{d+1}$ space are described by the
$so(d,2)$ algebra. Global symmetries of anomalous conformal currents and
shadow fields in $d$-dimensional space are also described by the $so(d,2)$
algebra. We have discussed global symmetries of anomalous conformal currents
and shadow fields by using conformal basis of the $so(d,2)$ algebra in
\rf{man01-12112010-01}. Therefore, for the studying AdS/CFT correspondence,
it is convenient to realize the relativistic symmetries of fields in
$AdS_{d+1}$ space by using also the conformal basis of the $so(d,2)$ algebra.
To achieve the conformal basis realization of bulk $so(d,2)$ symmetries we
use the Poincar\'e parametrization of AdS space \rf{lineelem01}.%
\footnote{ We note that, in our approach, only $so(d-1,1)$ symmetries are realized
manifestly. Symmetries of the $so(d,2)$ algebra could be realized manifestly by using the
framework of ambient space approach (see, e.g.,
Refs.\cite{Metsaev:1995re}-\cite{Bekaert:2009fg}).}
In the Poincar\'e parametrization, the $so(d,2)$ algebra transformations of arbitrary spin
massive AdS field $\phik$ is given by $\delta_{\hat{G}}\phik = \hat{G} \phik$, where the
realization of the $so(d,2)$ algebra generators $\hat{G}$ on space of $\phik$ takes the form
\beq
\label{conalggenlis01ads} && P^a = \partial^a \,,
\\[3pt]
\label{conalggenlis02ads} && J^{ab} = x^a\partial^b -  x^b\partial^a +
M^{ab}\,,
\\[3pt]
\label{conalggenlis03ads} && D = x\partial + \Delta\,, \qquad \Delta =
z\partial_z + \frac{d-1}{2}\,, \qquad\quad
\\[3pt]
\label{conalggenlis04ads} && K^a = K_{\Delta,M}^a + R^a\,,
\\[3pt]
&& \hspace{1cm} K_{\Delta,M}^a= -\half x^2\partial^a + x^a D + M^{ab}x^b
\,,\qquad
\\[3pt]
\label{14092008-06} &&  R^a  = R_\smzero^a + R_\smone^a\,,
\\[3pt]
\label{man01-26012012-15} && R_\smzero^a = z \Cwt^a ( r_\zeta \bar\zeta + r_z
\bar\alpha^z) - z (\zeta r_\zeta +\alpha^z r_z )\bar\alpha^a\,,
\\[5pt]
\label{14092008-08} && R_\smone^a =  -\half z^2 \partial^a\,,
\eeq
and the operator $\Cwt^a$ is given in \rf{man01-16102011-05}.

{\bf Modified de Donder gauge}. Gauge invariant equations of motion obtained
from Lagrangian \rf{man01-15102011-23} take the form
\beq \label{10072009-04} && \hspace{-1cm} \mubf \Box_\nu \phik - C \Cb \phik =0\,,
\\
&& \label{09072009-09}  \Box_\nu \equiv \Box + \partial_z^2 -
\frac{1}{z^2}(\nu^2 -\frac{1}{4})\,,
\\
\label{10072009-05}
&& C \equiv \alpar - \half \alpha^2 \albpar  - e_1 \Pi^\smponetwo + \half \eb_1
\alpha^2\,,\eeq
where $\Cb$ and $\nu$ are given in \rf{080405-01add} and
\rf{man01-16102011-03}. Note that, for the derivation of \rf{10072009-04}, we
use the relations $C^\dagger= - \Cb$,
\be  \label{10072009-06} \TT_{\nu-\half}^\dagger\TT_{\nu-\half} =
-\partial_z^2 + \frac{1}{z^2}(\nu^2-\frac{1}{4})\,.\ee
Modified de Donder gauge is defined to be
\be \label{man01-15102011-31} \Cb\phik = 0\,,\qquad \hbox{modified de Donder
gauge} \,,\ee
where $\Cb$ is given in \rf{080405-01add}. Using this gauge in
\rf{10072009-04} leads to the simple gauge-fixed equations of motion,
\beq \label{man01-15102011-27}
\Box_\nu |\phi\rangle  =0\,,
\eeq
i.e., the gauge-fixed equations turn out to be decoupled.

We note that the modified de Donder gauge and gauge-fixed equations have
on-shell leftover gauge symmetry. This is to say that modified de Donder
gauge \rf{man01-15102011-31} and gauge-fixed equations \rf{man01-15102011-27}
are invariant with respect to gauge transformation given in
\rf{man01-15102011-26} provided the gauge transformation parameter satisfies
the following equation:
\be \label{man01-15102011-33} \Box_\nu |\xi\rangle = 0\,.\ee

\subsection{ AdS/CFT correspondence for spin-0 field}

As we have already said, the use of our massive gauge fields and the modified de Donder gauge
makes the study of AdS/CFT correspondence for arbitrary spin-$s$ massive AdS field similar to
the one for spin-0 massive AdS field. Therefore, for the reader convenience, we now briefly
recall the AdS/CFT correspondence for the scalar massive AdS field.

{\bf AdS/CFT correspondence for normalizable
modes of spin-0 massive AdS field and spin-0 conformal current}%
\footnote{Also, see Ref.\cite{Balasubramanian:1998sn}.}.
The action of massive scalar field in $AdS_{d+1}$ background takes the form
\beq \label{19072009-01}
&& S =  \int d^dx dz\,  \LL \,,
\\
\label{19072009-02} && \LL = \half \sqrt{|g|}\left(g^{\mu\rho}\partial_\mu
\Phi
\partial_\rho \Phi + m^2 \Phi^2 \right)\,.  \eeq
Using the canonically normalized field $\phi$ defined by relation $\Phi=
z^{\frac{d-1}{2}}\phi$, we represent Lagrangian \rf{19072009-02} as (up to total derivative)
\beq
\label{19072009-04}
&& \LL =  \half |d\phi|^2 + \half |\TT_{\nu -\half } \phi|^2\,,
\\[5pt]
\label{09072009-04} && \hspace{1cm} \TT_\nu \equiv \partial_z +
\frac{\nu}{z}\,,
\qquad \nu = \sqrt{m^2 + \frac{d^2}{4}}\,.\qquad
\eeq
Note that only for spin-0 massive AdS field the $\nu$ takes the form given in
\rf{09072009-04}. For spin-$s$ massive AdS field with $s>0$ the $\nu$ is given in
\rf{man01-16102011-03}. The equation of motion obtained from Lagrangian \rf{19072009-04} is
given by
\be \label{19072009-06}
\Box_\nu  \phi = 0 \,,
\ee
where $\Box_\nu$ is defined in \rf{09072009-09}. The normalizable solution of
Eq.\rf{19072009-06} takes the form
\beq
\label{man01-16112010-01} && \phi(x,z) = U_\nu^\sc \phi_\cur(x)\,,
\\
\label{man01-01112010-31x} && \hspace{1cm} U_\nu^\sc \equiv h_\nu \sqrt{zq}
J_\nu(zq) q^{-(\nu + \half)}\,,
\\
&& \hspace{1cm} h_\nu\equiv 2^\nu\Gamma(\nu+1)\,,\qquad  q^2\equiv
\Box\,,\qquad \eeq
where $J_\nu$ stands for the Bessel function. The asymptotic behavior of
solution \rf{man01-16112010-01} is given by
\be \label{man01-16112010-02}
\phi(x,z) \ \ \stackrel{z\rightarrow 0}{\longrightarrow} \ \ z^{\nu + \half}
\phi_\cur(x)\,. \ee
From \rf{man01-16112010-02}, we see that the field $\phi_\cur$ is indeed the asymptotic
boundary value of the normalizable solution.

For the case of scalar field, there are no gauge symmetries and gauge
conditions to be matched. All that is needed to complete the AdS/CFT
correspondence is to match the bulk global symmetries of the AdS field
$\phi(x,z)$ and the respective boundary global symmetries of the current
$\phi_\cur(x)$. Realization of the global symmetries on AdS side and CFT side
is given in \rf{conalggenlis01ads}-\rf{14092008-08} and
\rf{conalggenlis01}-\rf{conalggenlis04} respectively. Obviously, the
Poincar\'e symmetries match automatically. Introducing the notation
$D_{_{\AdS}}$ and $D_{_{\CFT}}$ for the respective realizations of $D$
symmetry on bulk fields \rf{conalggenlis03ads} and boundary conformal
currents \rf{conalggenlis03}, we get the relation
\be \label{oldman-23122012-01}
D_{_{\AdS}} \phi(x,z) =  U_\nu^\sc D_{_{\CFT}} \phi_\cur(x)\,,
\ee
where $D_{_{\CFT}}$ corresponding to the conformal spin-0 current $\phi_\cur$
is obtained from \rf{conalggenlis03} by using $\Delta = \frac{d}{2}+\nu$ with
$\nu$ given in \rf{09072009-04}. From \rf{oldman-23122012-01}, we see that
$D$ symmetries of $\phi(x,z)$ and $\phi_\cur(x)$ also match. Finally, using
$R^a\phi_\cur(x)=0$ and taking into account that, for spin-0 AdS field,
$R_\smzero^a\phi(x,z)=0$, we see that the $K^a$ symmetries also match.

{\bf AdS/CFT correspondence for non-normalizable modes of spin-0 massive AdS field and spin-0
shadow field}. As shown in Ref.\cite{wit}, the non-normalizable solution of
Eq.\rf{19072009-06} with the Dirichlet problem corresponding to the boundary shadow field
$\phi_\sh(x)$ can be presented as
\beq \label{19072009-09}
\phi(x,z) & = & \sigma \int d^dy\, G_\nu (x-y,z) \phi_\sh(y)\,,
\\
\label{10072009-12} && G_\nu(x,z) = \frac{c_\nu z^{\nu+\half}}{ (z^2+
|x|^2)^{\nu + \frac{d}{2}} }\,,
\\
\label{10072009-13}&& \ \ c_\nu \equiv \frac{\Gamma(\nu+\frac{d}{2})}{
\pi^{d/2} \Gamma(\nu)} \,.
\eeq
For the later use, we introduce the normalization factor $\sigma$ in relation
\rf{19072009-09}. We recall that commonly used value of $\sigma$ is achieved
by setting $\sigma=1$. The asymptotic behaviors of Green function
\rf{10072009-12} and solution \rf{19072009-09} are given by,
\beq
\label{10072009-14}
&& G_\nu(x,z) \ \ \ \stackrel{z \rightarrow 0}{\longrightarrow} \ \ \ z^{-\nu
+ \half} \delta^d(x)\,,
\\[5pt]
\label{man01-02-21072009-14}
&& \phi(x,z) \,\,\, \stackrel{z\rightarrow 0 }{\longrightarrow}\,\,\, z^{-\nu
+ \half} \sigma \phi_\sh(x)\,.
\eeq
Relation \rf{man01-02-21072009-14} tells us that solution \rf{19072009-09} has indeed
asymptotic behavior corresponding to the shadow field.

Taking into account \rf{19072009-06} and \rf{19072009-01},\rf{19072009-04}, we find the
well-known expression for effective action,%
\footnote{ As usually, since solution of the Dirichlet problem
\rf{19072009-09} tends to zero as $z\rightarrow \infty$, we ignore
contribution to $S_\eff$ when $z=\infty$.}
\beq
\label{19072009-07} - S_\eff & = & \int d^dx\,  \LL_\eff\Bigr|_{z\rightarrow
0} \,,
\\[5pt]
\label{19072009-08} \LL_\eff & = & \half \phi \TT_{\nu -\half } \phi
\,,\qquad
\eeq
Using solution of the Dirichlet problem \rf{19072009-09} in
\rf{19072009-07}, \rf{19072009-08}, we get the effective action
\be \label{man01-02-21072009-15}
-S_\eff  = \nu c_\nu \sigma^2 \int d^dx_1d^dx_2
\frac{\phi_\sh(x_1)\phi_\sh(x_2)}{|x_{12}|^{2\nu + d}}\,.
\ee
Plugging the commonly used value of $\sigma$, $\sigma=1$, into
\rf{man01-02-21072009-15}, we get the properly normalized effective action
found in Refs.\cite{Gubser:1998bc,Freedman:1998tz}. An interesting novelty of
our computation of the effective action is that we use the Fourier transform of the Green
function. For the details of our computation, see Appendix C in
Ref.\cite{Metsaev:2009ym}.

\section{ AdS/CFT correspondence for normalizable
modes of massive AdS field and anomalous conformal current}\label{secAdS/CFTcur}

We now ready to consider the AdS/CFT correspondence for the spin-$s$ massive AdS field and
spin-$s$ anomalous conformal current. We begin with the discussion of the normalizable
solution of Eq.\rf{man01-15102011-27}. The normalizable solution of Eq.\rf{man01-15102011-27}
is given by
\beq \label{man01-15102011-28}
&& |\phi(x,z)\rangle = U_\nu |\phi_\cur(x)\rangle \,,
\\
\label{man01-15102011-29} && \hspace{1cm} U_\nu \equiv h_\kappa (-)^{N_z}
\sqrt{zq} J_\nu(zq) q^{-(\nu + \half)}\,,
\\
\label{man01-26012012-16} && \hspace{1cm} h_\kappa\equiv
2^\kappa\Gamma(\kappa+1)\,,\qquad  q^2\equiv \Box\,, \qquad
\eeq
where we do not show explicitly the dependence of $U_\nu$ on $z$, $q$, and
the parameter $\kappa$ defined in \rf{man01-11102011-04}. The asymptotic
behavior of solution \rf{man01-15102011-28} takes the form
\be \label{man01-15102011-30}
|\phi(x,z)\rangle \ \ \stackrel{z\rightarrow 0}{\longrightarrow} \ \ z^{\nu +
\half} \frac{2^\kappa\Gamma(\kappa+1)}{2^\nu\Gamma(\nu+1)}(-)^{N_z}|\phi_\cur(x)\rangle\,.
\ee
From \rf{man01-15102011-30}, we see that $|\phi_\cur\rangle$ is indeed boundary value of the
normalizable solution. In the right-hand side of relation \rf{man01-15102011-28}, we use the
notation $|\phi_\cur\rangle$ because we are going to demonstrate that this boundary value is
indeed the gauge field appearing in the gauge invariant formulation of the spin-$s$ anomalous
conformal current  in Sec.\ref{sec03}. Namely, we are going to prove the following
statements:

\noindent {\bf i}) For normalizable solution \rf{man01-15102011-28}, modified
de Donder gauge condition \rf{man01-15102011-31} leads to the differential
constraint of the spin-$s$ anomalous conformal
current given in \rf{man01-10112011-09}.

\noindent {\bf ii}) On-shell leftover gauge transformation
\rf{man01-15102011-26} of normalizable solution \rf{man01-15102011-28} leads
to the gauge transformation of the spin-$s$
anomalous conformal current given in \rf{man01-10112011-22}%
\footnote{ Note that gauge transformation given in \rf{man01-15102011-26}
is off-shell gauge transformation. On-shell leftover gauge transformation
is obtained from  gauge transformation \rf{man01-15102011-26} by using gauge transformation
parameter which satisfies equation \rf{man01-15102011-33}.}.

\noindent {\bf iii}) On-shell bulk $so(d,2)$ symmetries of the normalizable
solution \rf{man01-15102011-28} amount to boundary $so(d,2)$ conformal
symmetries of the spin-$s$ anomalous conformal current.

To prove these statements we use the following relations for
the operator $U_\nu$:
\beq
\label{man01-01112010-35} && \TT_{\nu-\half} U_\nu  = U_{\nu-1}\,,
\\[5pt]
\label{man01-01112010-36} && \TT_{-\nu-\half} U_\nu  = - U_{\nu+1}\Box\,,
\\[5pt]
\label{man01-01112010-37} && \TT_{-\nu+\half} (zU_\nu)  = - z U_{\nu+1}\Box +
2 U_\nu \,,
\\[5pt]
\label{man01-01112010-37a1} && \Box_\nu (z U_{\nu+1}) = 2 U_\nu\,,
\eeq
which, in turn, can be derived by using the following textbook identities
for the Bessel function:
\be \label{man01-01112010-38}
\TT_\nu J_{\nu } = J_{\nu-1}\,,
\qquad
\TT_{-\nu} J_{\nu } = - J_{\nu + 1}\,.
\ee

{\bf Matching of bulk modified de Donder gauge and boundary constraint}. We
now demonstrate how differential constraint for the anomalous conformal
current \rf{man01-10112011-09} is obtained from modified de Donder gauge
condition \rf{man01-15102011-31}. Using \rf{man01-01112010-35} and
\rf{man01-01112010-36}, we find the important relations
\be \label{man01-15102011-60} e_1 U_\nu = U_\nu e_{1,\cur}\,,  \qquad \eb_1
U_\nu = U_\nu \eb_{1,\cur}\,. \ee
Acting with operator $\Cb$ \rf{080405-01add} on solution $|\phi\rangle$
\rf{man01-15102011-28} and using \rf{man01-15102011-60}, we obtain the
relation
\be \label{man01-15102011-34}
\Cb|\phi(x,z)\rangle = U_\nu \Cb_\cur |\phi_\cur(x)\rangle\,,
\ee
where $\Cb_\cur$ is given in \rf{man01-10112011-10}. From
\rf{man01-15102011-34}, we see that our modified de Donder gauge condition
\rf{man01-15102011-31} leads indeed to the differential constraint for the anomalous
conformal current given in \rf{man01-10112011-09}.

{\bf Matching of bulk and boundary gauge symmetries}. We now show how gauge transformation of
the anomalous conformal current \rf{man01-10112011-22} is related to the on-shell leftover
gauge transformation of the massive AdS field \rf{man01-15102011-26}. To this end we note
that on-shell leftover gauge transformation of massive AdS field is obtained from
\rf{man01-15102011-26} by plugging gauge transformation parameter that satisfies equation
\rf{man01-15102011-33} into \rf{man01-15102011-26}. The normalizable solution of equation for
the gauge transformation parameter \rf{man01-15102011-33} takes the form
\be
\label{man01-15102011-35}  |\xi(x,z)\rangle = U_\nu |\xi_\cur(x)\rangle\,,
\ee
where $U_\nu$ is given in \rf{man01-15102011-29}. On the one hand, plugging
\rf{man01-15102011-35} into \rf{man01-15102011-26} and using
\rf{man01-15102011-60}, we find that bulk on-shell leftover gauge
transformation takes the form
\be \label{man01-15102011-62} \delta |\phi(x,z)\rangle = U_\nu G_\cur|\xi_\cur(x)\rangle
\,.\ee
On the other hand, relation \rf{man01-15102011-28} leads to
\be \label{man01-15102011-63}
\delta|\phi(x,z)\rangle = U_\nu \delta|\phi_\cur(x)\rangle \,.\ee
Comparing \rf{man01-15102011-62} and \rf{man01-15102011-63}, we conclude that
bulk and boundary gauge symmetries does indeed match.

{\bf Matching of bulk and boundary global symmetries}. Now we are going to demonstrate the
matching of the $so(d,2)$ algebra generators for bulk massive AdS field in
\rf{conalggenlis01ads}-\rf{conalggenlis04ads} and the ones for the boundary anomalous
conformal current in \rf{conalggenlis01}-\rf{conalggenlis04}. Representation for generators
of the $so(d,2)$ algebra given in \rf{conalggenlis01ads}-\rf{conalggenlis04ads} is valid for
the gauge invariant theory of AdS fields. Note however that the modified de Donder gauge
respects the Poincar\'e and dilatation symmetries, but breaks the conformal boost symmetries
($K^a$ symmetries). This implies that realization for generators $P^a$, $J^{ab}$ and $D$
given in \rf{conalggenlis01ads}-\rf{conalggenlis03ads} is still valid for the gauge-fixed AdS
fields, while realization for the conformal boost generator $K^a$ given in
\rf{conalggenlis04ads} must be modified to restore $K^a$ symmetries of the gauge-fixed AdS
fields. We begin with matching of the Poincar\'e and dilatation symmetries. Matching of the
Poincar\'e symmetries is obvious: from \rf{conalggenlis01},\rf{conalggenlis02} and
\rf{conalggenlis01ads},\rf{conalggenlis02ads}, we see that bulk and boundary generators of
the Poincar\'e algebra, $P^a$, $J^{ab}$, coincide. Next, we consider the dilatation generator
$D$. To match the dilatation symmetries we need an explicit form of the solution for AdS
field equations of motion in \rf{man01-15102011-28}. Using the notation $D_{_{\AdS}}$ and
$D_{_{\CFT}}$ for the respective bulk dilatation generator in \rf{conalggenlis03ads} and
boundary dilatation generator in \rf{conalggenlis03}, we get the relation
\be \label{manold-24122012-01}
D_{_{\AdS}} |\phi(x,z)\rangle =  U_\nu D_{_{\CFT}} |\phi_\cur(x)\rangle\,,
\ee
where $D_{_{\CFT}}$ corresponding to $|\phi_\cur\rangle$ is obtainable from
\rf{conalggenlis03} and the conformal dimension operator given in \rf{man01-15102011-36a1}.
From \rf{manold-24122012-01}, we see that the generators $D_{_{\AdS}}$ and $D_{_{\CFT}}$
match.

We now match the $K^a$ symmetries. As we noted above, our modified de Donder gauge breaks the
$K^a$ symmetries. This can be seen as follows. Using realization of $K^a$ transformations in
\rf{conalggenlis04ads}, we find that the gauge-fixed massive AdS field satisfies the relation
\be \label{man01-15102011-36}  \Cb K^a|\phi\rangle  = -
2\Cb_\perp^a|\phi\rangle\,,
\ee
which tells us  that the modified de Donder gauge condition, $\Cb\phik=0$, is not invariant
under $K^a$ symmetries that are described by generator $K^a$ in \rf{conalggenlis04ads}.
Therefore, to restore the $K^a$ symmetries of the gauge-fixed AdS field theory, we should
modify the generator $K^a$ in \rf{conalggenlis04ads}. We modify the generator $K^a$ by using
the standard procedure. Namely, we add compensating gauge transformations to maintain the
$K^a$ symmetries. This is to say that, in order to find improved $K_\impr^a$ transformations
we start with the generic global $K^a$ transformations \rf{conalggenlis04ads} supplemented by
the suitable compensating gauge transformation
\be \label{man01-15102011-37}
K_\impr^a|\phi\rangle = K^a|\phi\rangle + G|\xi^{K^a}\rangle\,,
\ee
where $G$ is given in \rf{man01-15102011-26a1} and $|\xi^{K^a}\rangle$ stands for the
parameter of the compensating gauge transformation. Using the relation
\be \label{man01-15102011-38}  \Cb G |\xi^{K^a}\rangle = \Box_\nu
|\xi^{K^a}\rangle \ee
and \rf{man01-15102011-36}, we find
\be   \Cb K_\impr^a \phik  = - 2\Cb_\perp^a \phik +
\Box_\nu |\xi^{K^a}\rangle \,.
\ee
Requiring the improved $K_\impr^a$ transformations to maintain the modified
de Donder gauge condition,
\be \label{man01-15102011-39} \Cb K_\impr^a \phik = 0\,, \ee
we get the equation for $|\xi^{K^a}\rangle$,
\be \label{man01-15102011-40} \Box_\nu |\xi^{K^a}\rangle - 2\Cb_\perp^a \phik
=0 \,.\ee
From \rf{man01-15102011-40}, we see that the compensating gauge
transformation parameter $|\xi^{K^a}\rangle$ should satisfy the
nonhomogeneous second-order differential equation. Plugging normalizable
solution of equation \rf{man01-15102011-28} into \rf{man01-15102011-40}, we see
that equation \rf{man01-15102011-40} leads to the equation
\be \label{man01-15102011-41}  \Box_\nu |\xi^{K^a}(x,z)\rangle = 2U_\nu
\Cb_\perp^a |\phi_\cur(x)\rangle\,.\ee
Using identity \rf{man01-01112010-37a1}, the solution to Eq.\rf{man01-15102011-41} is
found to be
\be
\label{man01-15102011-42} |\xi^{K^a}(x,z)\rangle = z
U_{\nu + 1} \Cb_\perp^a|\phi_\cur(x)\rangle\,. \ee
Plugging \rf{man01-15102011-28} and \rf{man01-15102011-42} into \rf{man01-15102011-37}, we
check that the improved $K_\impr^a$ transformations of spin-$s$ massive AdS field lead to the
conformal boost transformations of the spin-$s$ anomalous conformal current given in
\rf{04092008-01},\rf{conalggenlis04} with operator $R_\cur^a$ defined in
\rf{man01-15102011-43} (for details, see Appendix \ref{appen01}).

\section{ AdS/CFT correspondence for non-normalizable
modes of massive AdS field and anomalous shadow field}\label{secAdS/CFTsh}

We now discuss the AdS/CFT correspondence for bulk spin-$s$ massive AdS field and boundary
spin-$s$ anomalous shadow field. We begin with an analysis of the non-normalizable solution
of Eq.\rf{man01-15102011-27}. Because gauge-fixed equation of motion \rf{man01-15102011-27}
is similar to the one for the scalar AdS field \rf{19072009-06} we can simply apply result in
Sec. \ref{secAdS/CFT}. This is to say that solution of Eq.\rf{man01-15102011-27} with the
Dirichlet problem corresponding to the spin-$s$ anomalous shadow field takes the form
\beq \label{man01-15102011-47}
|\phi(x,z)\rangle  & = &  \sigma_\nu \int d^dy\, G_\nu (x-y,z)
|\phi_\sh(y)\rangle\,,
\\
\label{10012011-19} && \sigma_\nu \equiv \frac{2^\nu\Gamma(\nu)}{
2^\kappa\Gamma(\kappa)}(-)^{N_z} \,,
\eeq
where the Green function is given in \rf{10072009-12}.

Using asymptotic behavior of the Green function $G_\nu$ \rf{10072009-14}, we
find the asymptotic behavior of our solution
\be \label{man01-15102011-48}
|\phi(x,z)\rangle  \,\,\, \stackrel{z\rightarrow 0 }{\longrightarrow}\,\,\,
z^{-\nu + \half} \sigma_\nu |\phi_\sh(x)\rangle\,.
\ee
From this expression, we see that solution \rf{man01-15102011-47} has indeed
asymptotic behavior corresponding to the spin-$s$ anomalous shadow field.%
\footnote{ Since solution \rf{man01-15102011-47} has nonintegrable asymptotic
behavior \rf{man01-15102011-48}, such solution is sometimes referred to as
the non-normalizable solution.}
In the right-hand side of \rf{man01-15102011-47} we use the notation
$|\phi_\sh\rangle$, because we are going to show that this boundary value is
indeed the gauge field entering the gauge invariant formulation of the
spin-$s$ anomalous shadow field in Sec.\ref{sec04}. Namely, we are going to prove the
following statements:

\noindent {\bf i}) For solution \rf{man01-15102011-47}, modified
de Donder gauge condition \rf{man01-15102011-31} leads to differential
constraint of the anomalous shadow field given in \rf{man01-10112011-09sh}.

\noindent {\bf ii}) On-shell leftover gauge transformation
\rf{man01-15102011-26} of solution \rf{man01-15102011-47} leads to gauge
transformation of the spin-$s$
anomalous shadow field given in \rf{man01-10112011-22sh}%

\noindent {\bf iii}) On-shell global $so(d,2)$ symmetries of solution
\rf{man01-15102011-47} become global $so(d,2)$ conformal symmetries of the
spin-$s$ anomalous shadow field.

\noindent {\bf iv}) an effective action evaluated on solution
\rf{man01-15102011-47}  coincides, up to normalization factor, with boundary
two-point gauge invariant vertex for the anomalous shadow field given in
\rf{10112011-03}.

Below we demonstrate how these statements can be proved by using the
following relations for the Green function $G_\nu \equiv G_\nu(x-y,z)$:
\beq
\label{man01-01112010-24} && \TT_{-\nu+\half}G_{\nu-1} = - 2(\nu-1) G_\nu\,,
\\
\label{man01-01112010-24a1} && \TT_{\nu+\half}G_{\nu+1} = \frac{1}{2\nu} \Box G_\nu\,,
\\
&& \label{2m26122010-04}  \Box_\nu (zG_{\nu-1}) = - 4(\nu-1) G_\nu \,.
\eeq

{\bf Matching of bulk modified de Donder gauge and boundary differential
constraint}. We note that it is choice of normalization factor $\sigma_\nu$
in \rf{10012011-19} that allows us to match bulk modified de Donder gauge and
boundary differential constraint. Namely, the normalization factor
$\sigma_\nu$ \rf{10012011-19} is uniquely determined by the following two
requirements:

\noindent {\bf i}) The factor $\sigma_\nu$ is normalized to be
\be \label{man01-15102011-49} \sigma_\nu = 1\,,  \quad \hbox{ for } \quad N_z =0
\,, \quad N_\zeta = 0\,. \ee

\noindent {\bf ii}) Modified de Donder gauge condition for AdS field $\phik$
\rf{man01-15102011-31} should lead to the differential constraint for the
shadow field $|\phi_\sh\rangle$ \rf{man01-10112011-09sh}.

We note that the choice of normalization condition \rf{man01-15102011-49} is
a matter of convenience. This commonly used condition implies the following
normalization of asymptotic behavior of solution \rf{man01-15102011-47} for
the leading rank-$s$ tensor field $\phi_0^{a_1\ldots a_s}$ in
\rf{man01-15102011-19}:
\be \phi_0^{a_1\ldots a_s}(x,z)\,\,\,\stackrel{z \rightarrow 0}{\longrightarrow}\,\,\,
z^{-\kappa + \half} \phi_{\sh,0}^{a_1\ldots a_s}(x)\,, \ee
where $\phi_{\sh,0}^{a_1\ldots a_s}$ is the leading rank-$s$ tensor field in
\rf{man01-10112011-01sh}.

Using \rf{man01-01112010-24},\rf{man01-01112010-24a1} we find the relations
\be \label{man01-16102011-01} e_1 \sigma_\nu G_\nu = \sigma_\nu G_\nu
e_{1,\sh}\,, \qquad \eb_1 \sigma_\nu G_\nu = \sigma_\nu G_\nu
\eb_{1,\sh}\,,\ee
where Laplace operator $\Box$ appearing in $e_{1,\sh}$, $\eb_{1,\sh}$ is
acting on the Green function $G_\nu$. Acting with operator $\Cb$
\rf{080405-01add} on solution \rf{man01-15102011-47} and using
\rf{man01-16102011-01}, we find the relation
\be \label{man01-15102011-50}
\Cb |\phi\rangle  =  \sigma_\nu \int d^dy\, G_\nu(x-y,z)\Cb_\sh
|\phi_\sh(y)\rangle\,.
\ee
From this relation, we see that
modified de Donder gauge for the spin-$s$ massive AdS field \rf{man01-15102011-31} and
differential constraint for the spin-$s$ anomalous shadow field given in
\rf{man01-10112011-09sh} match.

{\bf Matching of bulk and boundary gauge symmetries}. We now show how gauge transformation of
the anomalous shadow field \rf{man01-10112011-22sh} is obtained from the on-shell leftover
gauge transformation of the massive AdS field \rf{man01-15102011-26}. To this end we note
that the corresponding on-shell leftover gauge transformation of massive AdS field is
obtained from \rf{man01-15102011-26} by plugging non-normalizable solution of equation for
gauge transformation parameter \rf{man01-15102011-33} into \rf{man01-15102011-26}. The
non-normalizable solution of equation \rf{man01-15102011-33} is given by
\be \label{man01-15102011-56} |\xi(x,z)\rangle  =  \sigma_\nu \int d^dy\,
G_\nu(x-y,z) |\xi_\sh(y)\rangle\,, \ee
where $\sigma_\nu$ is given in \rf{10012011-19}. We now note that, on the one
hand, plugging \rf{man01-15102011-56} into \rf{man01-15102011-26} and using
\rf{man01-16102011-01}, we can cast the on-shell leftover gauge
transformation of $|\phi\rangle$ into the form
\be \label{man01-15102011-57}
\delta |\phi\rangle  =  \sigma_\nu\int d^dy\, G_\nu(x-y,z) G_\sh
|\xi_\sh(y)\rangle\,.
\ee
On the other hand, making use of relation \rf{man01-15102011-47}, we get
\be \label{man01-15102011-58}
\delta \phik   =  \sigma_\nu \int d^dy\, G_\nu (x-y,z) \delta
|\phi_\sh(y)\rangle\,.
\ee
Comparing \rf{man01-15102011-57} with \rf{man01-15102011-58}, we conclude that the on-shell
leftover gauge symmetries of solution of the Dirichlet problem for the spin-$s$ massive AdS
field are indeed related to gauge symmetries of the spin-$s$ anomalous shadow field.

{\bf Matching of bulk and boundary global symmetries}. Matching of global symmetries can be
demonstrated by using the procedure we described for the spin-$s$ anomalous conformal current
in Sec.\ref{secAdS/CFTcur}. Therefore, to avoid repetition, let us briefly discuss only
relevant details. The matching of bulk and boundary symmetries of the Poincar\'e algebra is
straightforward. To match bulk and boundary dilatation symmetries all that we needed are
solution for bulk field in \rf{man01-15102011-47}, bulk dilatation operator
\rf{conalggenlis03ads}, and conformal dimension operator for the spin-$s$ anomalous shadow
field given in \rf{man01-15102011-59x1}. To match conformal boost symmetries we introduce
improved bulk $K_\impr^a$ transformations with compensating gauge transformation parameter
that satisfies Eq.\rf{man01-15102011-40} with $|\phi\rangle$ as in \rf{man01-15102011-47}.
Using relation \rf{2m26122010-04}, we find that the solution to Eq.\rf{man01-15102011-40}
with $|\phi\rangle$ as in \rf{man01-15102011-47} is given by
\be \label{man01-15102011-59}
|\xi^{K^a}(x,z)\rangle =  - z \sigma_{\nu-1}\! \int\!\! d^dy\, G_{\nu-1}(x-y,z)
\Cb_\perp^a |\phi_\sh(y)\rangle\,.
\ee
Using \rf{man01-15102011-47} and \rf{man01-15102011-59}, we check that
improved bulk $K_\impr^a$ transformations lead to $K^a$ transformations of
the spin-$s$ anomalous shadow field given in \rf{conalggenlis04} and
\rf{man01-15102011-61}.

{\bf Matching of effective action and boundary two-point vertex}. To find the
effective action we follow the standard procedure. Namely, we plug
non-normalizable solution of the bulk equation of motion with the Dirichlet
problem corresponding to the boundary anomalous shadow field
\rf{man01-15102011-47} into the bulk action \rf{man01-15102011-23a1}. We
proceed as follows. Using gauge invariant equation of motion \rf{10072009-04}
in \rf{man01-15102011-23a1}, we get the following expression for the
effective action:
\beq
\label{man01-15102011-44} S_\eff & = & - \int d^dx\,
\LL_\eff\Bigr|_{z\rightarrow 0} \,,
\\[5pt]
\label{man01-15102011-45} \LL_\eff & = & \half \langle \phi| \mubf \TT_{\nu
-\half } |\phi\rangle + \half \langle\phi| Y \Cb|\phi\rangle\,,
\nonumber\\
&& Y\equiv \half \alpha^2 (r_\zeta\bar\zeta + r_z\bar\alpha^z )- \zeta
r_\zeta - \alpha^z r_z \,.\qquad
\eeq
From \rf{man01-15102011-45}, it is clear that the use of modified de
Donder gauge condition \rf{man01-15102011-31} considerably simplifies the
expression for $\LL_\eff$,
\be \label{man01-15102011-46}
\LL_\eff\Bigr|_{\Cb\phik=0} =  \half \langle \phi|\mubf \TT_{\nu -\half }
\phik\,.
\ee
Thus we see that it is the use of modified de Donder gauge that leads to
$\LL_\eff$ for massive arbitrary spin AdS field \rf{man01-15102011-46} which
has the same form as $\LL_\eff$ for scalar field \rf{19072009-08}. Therefore,
to find $S_\eff$ for the massive arbitrary spin AdS field we can use results
for the scalar field. Namely, all that remains to obtain the effective action
is to plug solution \rf{man01-15102011-47} into \rf{man01-15102011-44},
\rf{man01-15102011-46}
and use general formula given in \rf{man01-02-21072009-15}. Doing so, we get%
\footnote{ All that is needed for the derivation of \rf{man01-15102011-51} is
to make the replacement $\sigma\rightarrow \sigma_\nu$ in formula for scalar
field \rf{man01-02-21072009-15} and note the easily derived algebraic
relation $\nu c_\nu \sigma_\nu^2 = \kappa c_\kappa f_\nu$, where $f_\nu$ is
defined in \rf{man01-15102011-11}.}
\be \label{man01-15102011-51}
-S_\eff  =  2\kappa c_\kappa \Gamma \,,
\ee
where $\kappa$ and $c_\kappa$ are given in
\rf{man01-11102011-04},\rf{10072009-13}, respectively, while $\Gamma$ stands
for gauge invariant two-point vertex of the spin-$s$ anomalous shadow field
given in \rf{10112011-03},\rf{man01-14102011-02}.

To summarize, using the modified de Donder gauge for the computation of the
spin-$s$ massive AdS field action on the solution of equations of motion with
the Dirichlet problem corresponding to the boundary anomalous shadow field,
we obtain the gauge invariant two-point vertex of the spin-$s$ anomalous
shadow field. It is the matching of the bulk leftover on-shell gauge
symmetries of solution to the Dirichlet problem and bulk global symmetries
and the respective boundary gauge symmetries of the anomalous shadow field
and boundary global symmetries that explains why the effective action of the
AdS massive field coincides, up to a normalization factor, with the gauge
invariant two-point vertex for the boundary anomalous shadow field.

In the literature, the effective action is expressed in terms of the
two-point vertex taken in the Stueckelberg gauge frame $\Gamma^\stand$
\rf{man01-15102011-53}. To express our result in terms of $\Gamma^\stand$ ,
we use relations \rf{man01-14102011-03},\rf{man01-15102011-54} to represent
our result \rf{man01-15102011-51} as
\be \label{man01-15102011-52}
-S_\eff  =  \frac{\kappa(2\kappa+2s+d-2)}{s!(2\kappa+d-2)} c_\kappa \Gamma^{{\rm stand}} \,.
\ee
The following remarks are in order:
\\
{\bf i}) For the particular values $s=1$ and $s=2$, our results for
normalization factor in front of $\Gamma^{{\rm stand}}$
\rf{man01-15102011-52} coincide with the ones obtained in
Refs.\cite{Mueck:1998iz} and \cite{Polishchuk:1999nh} respectively. Thus, our
results agree with the previously reported results for the particular values
$s=1,2$ and give the normalization factor for arbitrary values of $s$.

\noindent {\bf ii}) Using \rf{man01-15102011-52} and taking into account the
expression for $\Gamma^\stand$ given in
\rf{10112011-03},\rf{man01-15102011-53}, we see that the AdS/CFT
correspondence for massive spin-$s$ AdS field leads to two-point correlation
function of spin-$s$ anomalous conformal current with the conformal dimension
given by
\be \label{man01-16102011-12} \Delta = \frac{d}{2} + \sqrt{m^2 +
\Bigl(s+\frac{d-4}{2}\Bigr)^2} \,.\ee
According to the AdS/CFT correspondence this conformal dimension should be
equal to lowest energy value $E_0$ for massive spin-$s$ AdS field with mass
parameter $m$. Comparing $\Delta$ \rf{man01-16102011-12} with $E_0$ found in
Ref.\cite{Metsaev:2003cu} (see formula 5.74 in Ref.\cite{Metsaev:2003cu}) we
see that $E_0=\Delta$.

\noindent {\bf iii}) The fact that the effective action of massive AdS field is proportional
to the two-point vertex of anomalous shadow field is expected because of the conformal
symmetry. Note however that for the systematical study of AdS/CFT correspondence it is
important to know the normalization factor in front of $\Gamma^{{\rm stand}}$
\rf{man01-15102011-52}.

\noindent {\bf iv}) In the massless limit, $m\rightarrow 0$, our result for
$S_\eff$ \rf{man01-15102011-52} agrees with the previously reported results
in literature. Computation of $S_\eff$ for spin-1 and spin-2 massless fields
may be found in the respective Ref.\cite{Freedman:1998tz} and
Refs.\cite{Liu:1998bu}. Computation of $S_\eff$ for arbitrary spin-$s$
massless fields may be found in Ref.\cite{Metsaev:2009ym}. The study of
AdS/CFT correspondence for massless fields in light-cone gauge frame may be
found in Refs.\cite{Metsaev:1999ui}-\cite{Metsaev:2005ws} (see also
Ref.\cite{Koch:2010cy}).

\noindent {\bf v}) The effective action given in \rf{man01-15102011-51} is gauge invariant,
while the effective action given in \rf{man01-15102011-52} is obtained from one in
\rf{man01-15102011-51} by using the Stueckelberg gauge frame. One of advantages of our
approach is that our approach gives the possibility to study the effective action by using
other gauge conditions which might be preferable in various applications. For example, in the
light-cone gauge frame, the effective action given in \rf{man01-15102011-51} takes the form
\be \label{man01-15102011-55}
-S_\eff  =  2\kappa c_\kappa \Gamma^{({\rm l.c.})} \,,
\ee
where light-cone gauge vertex $\Gamma^{({\rm l.c.})}$ is given in
\rf{10112011-03},\rf{man01-15102011-10}. It is the formula
\rf{man01-15102011-55} that seems to be interesting  for the studying the
duality of light-cone gauge type IIB Green-Schwarz AdS superstring and
the corresponding CFT.

\section{Conclusions}\label{conlus}

In the present paper, we extended  our gauge invariant approach to CFT
initiated in Ref.\cite{Metsaev:2008fs} to the studying of arbitrary spin
anomalous conformal currents and shadow fields. We recall that, in the
framework of string/gauge theory duality, the anomalous conformal currents
and shadow fields are related to massive fields of AdS string theory. We note
that all Lorentz covariant approaches to string field theory involve large
amount of Stueckelberg fields (see, e.g., Ref.\cite{Siegel:1985tw}). Because
our approach to anomalous conformal currents and shadow fields also involves
Stueckelberg fields we believe that our approach will be helpful to
understand string/gauge theory duality better.

We obtained the gauge invariant vertex for anomalous shadow field which, in the framework of
AdS/CFT correspondence, is related to AdS field action evaluated on solution of the Dirichlet
problem. Our gauge invariant vertex provides quick and easy access to the light-cone gauge
vertex. Because one expects that the quantization of AdS superstring is straightforward only
in the light-cone gauge we believe that our light-cone gauge vertex will also be helpful in
various studies of AdS/CFT duality. Our results should have a number of the following
interesting applications and generalizations.

\noindent {\bf i}) In this paper, we studied bosonic anomalous conformal currents and shadow
fields. It would be interesting to extend our approach to the study of AdS/CFT correspondence
for arbitrary spin fermionic anomalous conformal currents and shadow fields and related
arbitrary spin massive fermionic fields \cite{Metsaev:2006zy}.

\noindent {\bf ii}) In this paper, we studied the AdS/CFT correspondence by
using CFT adapted approach to massive AdS fields developed in
Ref.\cite{Metsaev:2009hp}. In the last years, new interesting approaches to
massive AdS fields were developed (see, e.g., Refs.\cite{Ponomarev:2010st}).
It would be interesting to apply these new approaches to the study of AdS/CFT
correspondence for massive AdS fields.

\noindent {\bf iii}) An extension of our approach to the case of 3-point and 4-point gauge
invariant vertices of anomalous shadow fields will give us the possibility to study various
applications of our approach along the lines of Refs.\cite{Roiban:2010fe}.

\noindent {\bf iv}) Idea of arranging $d$-dimensional conformal physics in
$d+2$ dimensional multiplets was extensively studied in
Refs.\cite{Bars:2010xi}. Obviously, the use of the methods developed in
Refs.\cite{Bars:2010xi} may be very useful for the studying AdS/CFT
correspondence.

\noindent {\bf v}) The Becchi-Rouet-Stora-Tyutin (BRST) approach is one of
powerful methods of modern quantum field theory (see, e.g.,
Refs.\cite{Siegel:1999ew}). Obviously, an extension of BRST approach to the
case of anomalous conformal currents and shadow fields will provide new
interesting possibilities for the studying CFT.

\noindent {\bf vi}) Mixed-symmetry fields have extensively been studied in the last years
(see, e.g., Refs.\cite{Zinoviev:2008ve}). Needless to say that generalization of our approach
to the mixed-symmetry conformal currents and shadow fields could be of some interest.

\begin{acknowledgments}
This work was supported by RFBR Grant No.11-02-00685 and by the Alexander von Humboldt
Foundation Grant PHYS0167.
\end{acknowledgments}

\appendix
\section{ Matching of bulk and boundary conformal boost symmetries }
\label{appen01}

In this appendix, we demonstrate matching of the improved $K_\impr^a$ transformations of the
normalizable modes of massive AdS field \rf{man01-15102011-37} and the $K^a$ transformations
of boundary anomalous conformal current given in \rf{04092008-01},\rf{conalggenlis04} with
operator $R_\cur^a$ defined in \rf{man01-15102011-43}. Matching of improved $K_\impr^a$
transformations of non-normalizable modes of massive AdS field and boundary $K^a$
transformations of anomalous shadow fields can be demonstrated in a quite similar way.

We start with the realization of the improved $K_\impr^a$ transformations on
space of normalizable modes of massive gauge-fixed AdS field given by (see
\rf{man01-15102011-37})
\be \label{26012012-01} K_\impr^a|\phi_\norm\rangle  =
K_{_{\AdS}}^a|\phi_\norm\rangle + G_{_{\AdS}} |\xi_\norm^{K^a}\rangle\,, \ee
where, in this Appendix, the normalizable solution $\phik$ in \rf{man01-15102011-28} is
denoted by $|\phi_\norm\rangle$, while the normalizable solution for the compensating gauge
transformation parameter $|\xi^{K^a}\rangle$ in \rf{man01-15102011-42} is denoted by
$|\xi_\norm^{K^a}\rangle$. Also, in this Appendix, the generic generator of $K^a$ symmetries
in \rf{conalggenlis04ads} is denoted by $K_{_{\AdS}}^a$ , while the gauge transformation
operator $G$ in \rf{man01-15102011-26a1} is denoted by $G_{_{\AdS}}$. Our purpose is to
demonstrate that the improved $K_\impr^a$ transformations of the normalizable modes of
massive gauge-fixed AdS field become $K^a$ transformations of the anomalous conformal
current. Namely, we are going to prove the relation
\be \label{man01-26012012-03} K_\impr^a |\phi_\norm\rangle = U_\nu
K_{_{\CFT}}^a |\phi_\cur\rangle \,, \ee
where $K_{_{\CFT}}^a$ stands for the realization of the conformal boost
generator on space of the anomalous conformal current given in
\rf{conalggenlis04},\rf{man01-15102011-43}.

In order to prove relation \rf{man01-26012012-03} we represent the operator
$K_{_{\AdS}}^a$ \rf{conalggenlis04ads} as
\beq
\label{man01-26012012-04}  && K_{_{\AdS}}^a = K_{\Delta_{\AdS}}^a +
R_\smone^a + M^{ab} x^b + R_\smzero^a \,,
\\
&& \label{man01-26012012-05} \quad\qquad K_{\Delta_{\AdS}}^a \equiv -\half
x^2\partial^a + x^a D_{_{\AdS}} \,, \eeq
where operators $D_{_{\AdS}}$, $R_\smzero^a$, and $R_\smone^a$ are given in
\rf{conalggenlis03ads}, \rf{man01-26012012-15}, and \rf{14092008-08} respectively. Next, we
note the relations
\beq
\label{man01-26012012-09} && \hspace{-0.7cm}(K_{\Delta_{\AdS}}^a +
R_\smone^a)|\phi_\norm\rangle = U_\nu K_{\Delta_\cur}^a |\phi_\cur\rangle\,,
\qquad
\\[7pt]
\label{man01-26012012-10} && \hspace{-0.7cm}(M^{ab}x ^b + R_\smzero^a)|
\phi_\norm\rangle + G_{_{\AdS}}|\xi_\norm^{K^a} \rangle
\nonumber\\[5pt]
&& \hspace{1.7cm} = U_\nu ( M^{ab}x^b + R_\cur^a ) |\phi_\cur\rangle\,,
\eeq
where

\beq
&&  K_{\Delta_\cur}^a \equiv -\half x^2\partial^a + x^a D_\cur\,,
\\
&& D_\cur \equiv  x\partial + \Delta_\cur\,,
\eeq
while $\Delta_\cur$ and $R_\cur^a$ are given in \rf{man01-15102011-36a1} and
\rf{man01-15102011-43} respectively. Using \rf{man01-26012012-09},\rf{man01-26012012-10}, we
see that relation \rf{man01-26012012-03} does indeed hold.

We now comment on the derivation of relations
\rf{man01-26012012-09},\rf{man01-26012012-10}. These relations are obtained
by using the following general formulas

\beq
\label{man01-26012012-11} && (K_{\Delta_{\AdS}}^a + R_\smone^a) U_\nu = U_\nu
(K_{\Delta_\cur}^a + x^a z\partial_z )
\nonumber\\[5pt]
&& \hspace{1cm} - \,   h_\kappa(-)^{N_z}q^{-\nu - \frac{3}{2}} \partial^a (\partial_q Z_\nu
(qz))z\partial_z\,,
\\[5pt]
\label{man01-26012012-12} && (M^{ab}x ^b + R_\smzero^a) U_\nu +  G_{_{\AdS}}
(zU_{\nu +1} \Cb_\perp^a)
\nonumber\\[5pt]
&& \hspace{2.7cm} \approx U_\nu ( M^{ab}x^b + R_\cur^a )\,,
\eeq
where $Z_\nu(z) \equiv \sqrt{z} J_\nu(z)$, while  $q$ is defined in \rf{man01-26012012-16}.
In \rf{man01-26012012-12} and relation \rf{man01-26012012-17} given below, the sign $\approx$
implies that relations \rf{man01-26012012-12}, \rf{man01-26012012-17} are valid only on space
of the anomalous conformal current $|\phi_\cur\rangle$. Recall that $|\phi_\cur\rangle$ is
subject to differential constraint \rf{man01-10112011-09}. We now see that, by applying
relations \rf{man01-26012012-11} and \rf{man01-26012012-12} to the anomalous conformal
current $|\phi_\cur\rangle$, we obtain the respective relations \rf{man01-26012012-09} and
\rf{man01-26012012-10}.

For the reader convenience, we write down the helpful formulas to be used for
the derivation of relation \rf{man01-26012012-12},
\beq
\label{man01-26012012-17} && M^{ab}x ^b U_\nu \approx  U_\nu  M^{ab}x^b
\nonumber\\[5pt]
&& \qquad  +  z U_{\nu+1} ( - G_\cur \Cb_\perp^a - e_{1,\cur}\bar\alpha^a +
\Cwt^a \eb_{1,\cur})\,,\qquad
\\[5pt]
\label{nman01-26012012-18} && R_\smzero^a U_\nu =   zU_{\nu+1}(\alpha^z r_z
\bar\alpha^a + \Cwt^a r_\zeta\bar\zeta)
\nonumber\\
&& \qquad \qquad  - \, z U_{\nu - 1} (\zeta r_\zeta \bar\alpha^a + \Cwt^a
r_z\bar\alpha^z)\,,
\\[5pt]
\label{man01-26012012-19} && G_{_{\AdS}} (zU_{\nu +1} \Cb_\perp^a) =
zU_{\nu+1} G_\cur \Cb_\perp^a
\nonumber\\
&& \hspace{1cm} + \, U_\nu  (2 \zeta r_\zeta - 2 \alpha^2
\frac{1}{2N_\alpha+d-2} r_z \bar\alpha^z )\Cb_\perp^a \,, \qquad
\\
&& zU_{\nu -1} + z\Box U_{\nu+1}=2\nu U_\nu\,.
\eeq
In turn, relation \rf{man01-26012012-17} is obtained
by using the identity
\beq
&&  M^{ab} \partial^b + G_\cur \Cb_\perp^a
\nonumber\\[3pt]
&& \hspace{1cm} = \alpha^a \Cb_\cur - e_{1,\cur} \bar\alpha^a + \Cwt^a \eb_{1,\cur}
\eeq
and differential constraint \rf{man01-10112011-09}.


\small
\begin{thebibliography}{30}

\parskip=0.pt


\bibitem{Fradkin:1985am}
  E.~S.~Fradkin and A.~A.~Tseytlin,
  Phys.\ Rept.\  {\bf 119}, 233 (1985).
%
\\
%
  J.~Erdmenger,
  Class.\ Quant.\ Grav.\  {\bf 14}, 2061 (1997)
  [arXiv:hep-th/9704108].
%
\\
%
  N.~Boulanger and M.~Henneaux,
  Annalen Phys.\  {\bf 10}, 935 (2001)
  [arXiv:hep-th/0106065].
%
\\
%
  A.~Y.~Segal,
  Nucl.\ Phys.\ B {\bf 664}, 59 (2003)
  hep-th/0207212
%
\\
%
  O.~V.~Shaynkman, I.~Y.~Tipunin and M.~A.~Vasiliev,
  Rev.\ Math.\ Phys.\  {\bf 18}, 823 (2006)
  hep-th/0401086
%
\\
%
  M.~A.~Vasiliev,
  Nucl.\ Phys.\  B {\bf 829}, 176 (2010)
  [arXiv:0909.5226 [hep-th]].


\bibitem{Petkou:1994ad}
  A.~Petkou,
  Annals Phys.\  {\bf 249}, 180 (1996);
  [arXiv:hep-th/9410093].
%
   Phys.Lett.\ B {\bf 389}, 18 (1996)
  hep-th/9602054.


\bibitem{Metsaev:2008fs}
  R.~R.~Metsaev,
  Phys.\ Rev.\  D {\bf 78}, 106010 (2008)
  [arXiv:0805.3472 [hep-th]].


\bibitem{Metsaev:2009ym}
  R.~R.~Metsaev,
  Phys.\ Rev.\  D {\bf 81}, 106002 (2010)
  [arXiv:0907.4678 [hep-th]].


\bibitem{Metsaev:2010zu}
  R.~R.~Metsaev,
  Phys.\ Rev.\  D {\bf 83}, 106004 (2011)
  [arXiv:1011.4261 [hep-th]].


\bibitem{Maldacena:1997re}
  J.~M.~Maldacena,
  Adv.\ Theor.\ Math.\ Phys.\  {\bf 2}, 231 (1998)
  [Int.\ J.\ Theor.\ Phys.\  {\bf 38}, 1113 (1999)]
  [arXiv:hep-th/9711200].



\bibitem{Metsaev:1998it}
  R.~R.~Metsaev and A.~A.~Tseytlin,
  Nucl.\ Phys.\  B {\bf 533}, 109 (1998)
  [arXiv:hep-th/9805028].




\bibitem{Metsaev:2000yf}
  R.~R.~Metsaev and A.~A.~Tseytlin,
  Phys.\ Rev.\  D {\bf 63}, 046002 (2001)
  [arXiv:hep-th/0007036].



\bibitem{Metsaev:2000yu}
  R.~R.~Metsaev, C.~B.~Thorn and A.~A.~Tseytlin,
  Nucl.\ Phys.\  B {\bf 596}, 151 (2001)
  [arXiv:hep-th/0009171].


\bibitem{Metsaev:2000mv}
  R.~R.~Metsaev and A.~A.~Tseytlin,
  J.\ Math.\ Phys.\  {\bf 42}, 2987 (2001)
  [arXiv:hep-th/0011191].



\bibitem{Mueck:1998iz}
  W.~Mueck and K.~S.~Viswanathan,
  Phys.\ Rev.\  D {\bf 58}, 106006 (1998)
  [arXiv:hep-th/9805145].




\bibitem{Polishchuk:1999nh}
  A.~Polishchuk,
  JHEP {\bf 9907}, 007 (1999)
  hep-th/9905048.



\bibitem{Metsaev:2008ks}
  R.R. Metsaev,
  Phys. Lett. B {\bf 671}, 128 (2009)
  [arXiv:0808.3945]


\bibitem{Metsaev:2009hp}
  R.R. Metsaev,
  Phys. Lett. B {\bf 682}, 455 (2010)
  [arXiv:0907.2207]



\bibitem{Bekaert:2006ix}
  X.~Bekaert and N.~Boulanger,
  Commun.\ Math.\ Phys.\  {\bf 271}, 723 (2007)
  [arXiv:hep-th/0606198].

\bibitem{Boulanger:2008up}
  N.~Boulanger, C.~Iazeolla and P.~Sundell,
  JHEP {\bf 0907}, 013 (2009)
  [arXiv:0812.3615 [hep-th]].
%
  JHEP {\bf 0907}, 014 (2009)
  [arXiv:0812.4438 [hep-th]].



\bibitem{Metsaev:2010kp}
  R.~R.~Metsaev,
  J.\ Phys.\ A  {\bf 44}, 175402 (2011)
  [arXiv:1012.2079 [hep-th]].


\bibitem{Metsaev:2008ba}
  R.~R.~Metsaev,
  J.\ Phys.\ A  {\bf 43}, 115401 (2010)
  [arXiv:0812.2861 [hep-th]].



\bibitem{Metsaev:2007fq}
R.~R.~Metsaev,
JHEP {\bf 1201}, 064 (2012) [arXiv:0707.4437 [hep-th]];
%
``Ordinary-derivative formulation of conformal totally symmetric arbitrary
spin bosonic fields,'' arXiv:0709.4392 [hep-th].





\bibitem{Fronsdal:1978vb}
C.~Fronsdal,
Phys.\ Rev.\ D {\bf 20}, 848 (1979).




\bibitem{Francia:2002aa}
  D.~Francia and A.~Sagnotti,
  Phys.\ Lett.\ B {\bf 543}, 303 (2002)
  [arXiv:hep-th/0207002].
%
\\
%
  A.~Sagnotti and M.~Tsulaia,
  Nucl.\ Phys.\  B {\bf 682}, 83 (2004)
  [arXiv:hep-th/0311257].
%
\\
%
  I.~L.~Buchbinder, A.~V.~Galajinsky and V.~A.~Krykhtin,
  Nucl.\ Phys.\  B {\bf 779}, 155 (2007)
  [arXiv:hep-th/0702161].
%
\\
%
  A.~Campoleoni, D.~Francia, J.~Mourad and A.~Sagnotti,
  Nucl.\ Phys.\  B {\bf 815}, 289 (2009)
  [arXiv:0810.4350 [hep-th]].
%
\\
%
  A.~Fotopoulos and M.~Tsulaia,
  Int.\ J.\ Mod.\ Phys.\  A {\bf 24}, 1 (2009)
  [arXiv:0805.1346 [hep-th]].


\bibitem{Alkalaev:2003qv}
K.~B.~Alkalaev, O.~V.~Shaynkman and M.~A.~Vasiliev,
Nucl.\ Phys.\  B {\bf 692}, 363 (2004) [arXiv:hep-th/0311164].
%
%
arXiv:hep-th/0601225.



\bibitem{Gover:2008sw}
  A.~R.~Gover, A.~Shaukat and A.~Waldron,
  Nucl.\ Phys.\  B {\bf 812}, 424 (2009)
  [arXiv:0810.2867 [hep-th]].



\bibitem{Gover:2008pt}
  A.~R.~Gover, A.~Shaukat and A.~Waldron,
  Phys.\ Lett.\  B {\bf 675}, 93 (2009)
  [arXiv:0812.3364 [hep-th]].
%
\\
%
  A.~Shaukat and A.~Waldron,
  Nucl.\ Phys.\  B {\bf 829}, 28 (2010)
  [arXiv:0911.2477 [hep-th]].
%
\\
%
  A.~Shaukat,
  arXiv:1003.0534 [math-ph].



\bibitem{Gov}
%
T.N.~ Bailey, M.G.~Eastwood and A.R.~Gover, Rocky Mountain J. Math. 24, 1191
(1994).


\bibitem{Konstein:2000bi}
  S.E.Konstein, M.A.Vasiliev and V.N.Zaikin,
  JHEP {\bf 0012}, 018 (2000)
  arXiv:hep-th/0010239
%
\\
%
  O.~A.~Gelfond, E.~D.~Skvortsov and M.~A.~Vasiliev,
  Theor.\ Math.\ Phys.\  {\bf 154}, 294 (2008)
  [arXiv:hep-th/0601106].


\bibitem{Erdmenger:1997wy}
  J.~Erdmenger and H.~Osborn,
  Class.\ Quant.\ Grav.\  {\bf 15}, 273 (1998)
  [arXiv:gr-qc/9708040].


\bibitem{Guttenberg:2008qe}
S.~Guttenberg and G.~Savvidy,
SIGMAP bulletin 4, 061 (2008) arXiv:0804.0522 [hep-th].
%
\\
%
  R.~Manvelyan, K.~Mkrtchyan and W.~Ruhl,
  Nucl.\ Phys.\  B {\bf 803}, 405 (2008)
  [arXiv:0804.1211 [hep-th]].
%
\\
%
  A.~Fotopoulos and M.~Tsulaia,
  JHEP {\bf 0910}, 050 (2009)
  [arXiv:0907.4061 [hep-th]].




\bibitem{Chang:2011mz}
  C.~-M.~Chang, X.~Yin,
  [arXiv:1106.2580 [hep-th]].


\bibitem{Brodsky:2008pg}
  S.~J.~Brodsky and G.~F.~de Teramond,
  ``AdS/CFT and Light-Front QCD,''
  arXiv:0802.0514 [hep-ph].
%
  AIP Conf.\ Proc.\  {\bf 1257}, 59 (2010)
  [arXiv:1001.5193 [hep-ph]].
%
\\
%
  O.~Andreev,
  Phys.\ Rev.\  D {\bf 67}, 046001 (2003)
  hep-th/0209256;
%
  Phys.\ Rev.\  D {\bf 81}, 087901 (2010)
  [arXiv:1001.4414 [hep-ph]].



\bibitem{Zinoviev:2001dt}
  Yu.~M.~Zinoviev,
  ``On massive high spin particles in (A)dS,''
  arXiv:hep-th/0108192.



\bibitem{Bolotin:1999fa}
  K.~I.~Bolotin and M.~A.~Vasiliev,
  Phys.\ Lett.\  B {\bf 479}, 421 (2000)
  [arXiv:hep-th/0001031].
%
\\
%
  V.~E.~Didenko and M.~A.~Vasiliev,
  Phys.\ Lett.\  B {\bf 682}, 305 (2009)
  [arXiv:0906.3898 [hep-th]].



\bibitem{Metsaev:1995re}
  R.~R.~Metsaev,
  Phys.\ Lett.\ B {\bf 354}, 78 (1995).


\bibitem{Metsaev:1997nj}
  R.~R.~Metsaev, Lect. Notes Phys. {\bf 524}, 331 (1997)
  [arXiv:hep-th/9810231].


\bibitem{Bekaert:2009fg}
  X.~Bekaert and M.~Grigoriev,
  SIGMA {\bf 6}, 038 (2010)
  [arXiv:0907.3195 [hep-th]].
%
\\
%
  R.~Bonezzi, E.~Latini and A.~Waldron,
  Phys.\ Rev.\  D {\bf 82}, 064037 (2010)
  [arXiv:1007.1724 [hep-th]].
%
\\
%
  A.~Fotopoulos, K.~L.~Panigrahi and M.~Tsulaia,
  Phys.\ Rev.\  D {\bf 74}, 085029 (2006)
  [arXiv:hep-th/0607248].



\bibitem{Balasubramanian:1998sn}
  V.~Balasubramanian, P.~Kraus and A.~E.~Lawrence,
  Phys.Rev.D {\bf 59}, 046003 (1999)
  hep-th/9805171


\bibitem{wit}
E.Witten,\!
Adv.Theor.Math.Phys.\!  {\bf 2}, 253 (1998), hep-th/9802150


\bibitem{Gubser:1998bc}
  S.~S.~Gubser, I.~R.~Klebanov and A.~M.~Polyakov,
  Phys.\ Lett.\  B {\bf 428}, 105 (1998)
  [arXiv:hep-th/9802109].



\bibitem{Freedman:1998tz}
  D.~Z.~Freedman, S.~D.~Mathur, A.~Matusis and L.~Rastelli,
  Nucl.\ Phys.\  B {\bf 546}, 96 (1999)
  [arXiv:hep-th/9804058].


\bibitem{Metsaev:2003cu}
  R.~R.~Metsaev,
  Phys. Lett.  B {\bf 590}, 95 (2004)
  hep-th/0312297



\bibitem{Liu:1998bu}
  H.~Liu and A.~A.~Tseytlin,
  Nucl.\ Phys.\ B {\bf 533}, 88 (1998)
  [arXiv:hep-th/9804083].
%
\\
%
  G.~E.~Arutyunov and S.~A.~Frolov,
  Nucl.\ Phys.\  B {\bf 544}, 576 (1999)
  [arXiv:hep-th/9806216].
%
\\
%
  W.~Mueck and K.~S.~Viswanathan,
  arXiv:hep-th/9810151.



\bibitem{Metsaev:1999ui}
  R.~R.~Metsaev,
  Nucl.\ Phys.\ B {\bf 563}, 295 (1999)
  hep-th/9906217.



\bibitem{Metsaev:2002vr}
  R.~R.~Metsaev,
  Phys.\ Lett.\  {\bf B531}, 152-160 (2002).
  [hep-th/0201226].


\bibitem{Metsaev:2005ws}
  R.~R.~Metsaev,
  Phys.\ Lett.\  B {\bf 636}, 227 (2006)
  [arXiv:hep-th/0512330].


\bibitem{Koch:2010cy}
  R.~d.~M.~Koch, A.~Jevicki, K.~Jin and J.~P.~Rodrigues,
  Phys.\ Rev.\  D {\bf 83}, 025006 (2011)
  [arXiv:1008.0633 [hep-th]].





\bibitem{Siegel:1985tw}
  W.~Siegel and B.~Zwiebach,
  Nucl.\ Phys.\  B {\bf 263}, 105 (1986).



\bibitem{Metsaev:2006zy}
  R.~R.~Metsaev,
  Phys.\ Lett.\  B {\bf 643}, 205 (2006)
  [arXiv:hep-th/0609029].


\bibitem{Ponomarev:2010st}
  D.~S.~Ponomarev and M.~A.~Vasiliev,
  Nucl.\ Phys.\  B {\bf 839}, 466 (2010)
  [arXiv:1001.0062 [hep-th]].
%
\\
%
  M.~Grigoriev and A.~Waldron,
  Nucl.\ Phys.\  B {\bf 853}, 291 (2011)
  [arXiv:1104.4994 [hep-th]].
%
\\
%
  K.~Alkalaev and M.~Grigoriev,
  Nucl.\ Phys.\  B {\bf 853}, 663 (2011)
  [arXiv:1105.6111 [hep-th]].



\bibitem{Roiban:2010fe}
  R.~Roiban and A.~A.~Tseytlin,
  arXiv:1008.4921 [hep-th].
%
\\
%
  H.~Liu and A.~A.~Tseytlin,
  Phys.\ Rev.\  D {\bf 59}, 086002 (1999)
  [arXiv:hep-th/9807097].
%
\\
%
  T.~Leonhardt and W.~Ruhl,
  J.\ Phys.\ A  {\bf 36}, 1159 (2003)
  [arXiv:hep-th/0210195].
%
\\
%
  M.~S.~Costa, J.~Penedones, D.~Poland and S.~Rychkov,
  arXiv:1107.3554 [hep-th].


\bibitem{Bars:2010xi}
  I.~Bars,
  Int.\ J.\ Mod.\ Phys.\  A {\bf 25}, 5235 (2010)
  [arXiv:1004.0688 [hep-th]].
%
\\
%
  I.~Bars and S.~H.~Chen,
  Phys.\ Rev.\  D {\bf 79}, 085021 (2009)
  [arXiv:0811.2510 [hep-th]].
%
\\
%
  I.~Bars, C.~Deliduman and O.~Andreev,
  Phys.\ Rev.\  D {\bf 58}, 066004 (1998)
  [arXiv:hep-th/9803188].







\bibitem{Siegel:1999ew}
  W.~Siegel,
  ``Fields,''
  arXiv:hep-th/9912205.
%
\\
%
  I.~L.~Buchbinder and V.~A.~Krykhtin,
  Nucl.\ Phys.\ B {\bf 727}, 537 (2005)
  [arXiv:hep-th/0505092].
%
\\
%
  I.~L.~Buchbinder, V.~A.~Krykhtin and P.~M.~Lavrov,
  Nucl.\ Phys.\  B {\bf 762}, 344 (2007)
  hep-th/0608005
%
\\
%
  K.~B.~Alkalaev and M.~Grigoriev,
  Nucl.\ Phys.\  B {\bf 835}, 197 (2010)
  [arXiv:0910.2690 [hep-th]].


\bibitem{Zinoviev:2008ve}
  Yu.~M.~Zinoviev,
  Nucl.\ Phys.\  B {\bf 812}, 46 (2009)
  [arXiv:0809.3287 [hep-th]].
%
\\
%
  E.~D.~Skvortsov,
  JHEP {\bf 0807}, 004 (2008)
  [arXiv:0801.2268 [hep-th]].
%
  Nucl.\ Phys.\  B {\bf 808}, 569 (2009)
  [arXiv:0807.0903 [hep-th]].
%
  JHEP {\bf 1001}, 106 (2010)
  [arXiv:0910.3334 [hep-th]].
%
\\
%
  N.~Boulanger, E.~D.~Skvortsov,
  JHEP {\bf 1109}, 063 (2011).
  [arXiv:1107.5028 [hep-th]].
%
\\
%
  N.~Boulanger, E.~D.~Skvortsov, Y.~.M.~Zinoviev,
  J.\ Phys.\ A {\bf A44}, 415403 (2011).
  [arXiv:1107.1872 [hep-th]].
















\end{thebibliography}
\end{document}